\documentclass[12pt]{article}
\pdfoutput=1
\usepackage{appendix}
\usepackage{anysize}
\usepackage{graphics}
\usepackage{color}
\usepackage{amssymb}
\usepackage{graphicx}
\usepackage{epstopdf}
\usepackage[hang,small,bf]{caption}
\usepackage{hyperref}
\usepackage{amsmath}
\usepackage{latexsym}
\usepackage{setspace}
\usepackage{sectsty}
\usepackage{breqn}
\usepackage{appendix}

\usepackage{graphicx}
\usepackage{dcolumn}
\usepackage{bm}
\usepackage{epstopdf}
\usepackage{color}

\newcommand{\be}{\begin{equation}}
\newcommand{\ee}{\end{equation}}
\newcommand{\bea}{\begin{eqnarray}}
\newcommand{\eea}{\end{eqnarray}}

\def\4vol{{\int d^4x \sqrt{-g}}}

\def\simlt{\stackrel{<}{{}_\sim}}
\def\simgt{\stackrel{>}{{}_\sim}}
\def\beq{\begin{equation}}
\def\eeq{\end{equation}}
\def\bea{\begin{eqnarray}}
\def\eea{\end{eqnarray}}
\def\bitem{\begin{itemize}}
\def\eitem{\end{itemize}}

\newcommand{\nc}{\newcommand}
\nc{\nt}{\tilde{N}}
\nc{\ra}{\rightarrow}
\nc{\lsim}{\begin{array}{c}\,\sim\vspace{-21pt}\\< \end{array}}
\nc{\gsim}{\begin{array}{c}\sim\vspace{-21pt}\\> \end{array}}
\nc{\tnt}{\tilde{N}}
\nc{\tst}{\tilde{t}}

\nc{\LL}{L}
\nc{\vv}{\tilde{v}}
\title{
\begin{flushright}
\normalsize{
ANL-HEP-PR-12-27\\ EFI 12-7\\ FERMILAB-PUB-12-195-PPD-T
}
\end{flushright}
\vspace*{5mm} \Large\textbf{Light Stau Phenomenology  and the Higgs $\gamma\gamma$ Rate}
\vspace*{1.0cm}
\author{\textbf{Marcela Carena$^{a,b}$, Stefania Gori$^{a,c}$ } \\
\textbf{ Nausheen R.~Shah$^b$, Carlos E.~M.~Wagner$^{a,c,d}$, Lian-Tao Wang$^{a,d}$ } \\
~\\
\normalsize\emph{$^a$Enrico Fermi Institute, University of Chicago, Chicago, IL 60637} \\
\normalsize\emph{$^b$Fermi National Accelerator Laboratory, P.~O.~Box 500, Batavia, IL 60510\footnote{http://theory.fnal.gov}}\\
\normalsize\emph{$^c$HEP Division, Argonne National Laboratory, 9700 Cass Ave., Argonne, IL 60439}\\
\normalsize\emph{$^d$Kavli Institute for Cosmological Physics, University of Chicago, Chicago, IL 60637}
}
}
\begin{document}

\setcounter{page}{0}
\maketitle
\begin{abstract}
Recently, an excess of events consistent with a Higgs boson with mass of about 125 GeV was reported by the CMS and ATLAS experiments. This Higgs boson mass is consistent with the values that may be obtained in minimal supersymmetric extensions of the Standard Model (SM), with both stop masses less than a TeV and large mixing. The apparently enhanced photon production rate associated with this potential Higgs signal may be the result of light staus with large mixing. Large stau mixing and large coupling of the staus to the SM-like Higgs boson may be obtained for large values of $\tan \beta$  and moderate to large values of the Higgsino mass parameter, $\mu$. We study the phenomenological properties of this  scenario, including precision electroweak data, the muon anomalous magnetic moment, Dark Matter, and the evolution of the soft supersymmetry-breaking parameters to high energies. We also analyze the possible collider signatures of light third generation sleptons and demonstrate that  it is possible to find evidence of their production at the 8 TeV and the 14 TeV LHC. The most promising channel is stau and tau sneutrino associated production, with the sneutrino decaying into a $W$ boson plus a light stau. 
\end{abstract}
\thispagestyle{empty}
\newpage
\setcounter{page}{1}

\section{Introduction}
\label{sec:Intro}
Searches for the Standard Model~(SM) Higgs boson are ongoing at the Large Hadron Collider (LHC) at CERN, with an 8 TeV center of mass energy for the proton collisions, recently upgraded from 7~TeV.  The relatively modest amount of data accumulated in  2011 has already led to interesting bounds on  a SM-like Higgs well beyond those from LEP and from the Tevatron.  Higgs boson masses  in the (129-539) GeV range are excluded at the 95\% C.L. both by ATLAS~\cite{ATLAS:2012ae,AtlasTwiki} and CMS~\cite{Chatrchyan:2012tx,CMSTwiki}, unless there is new physics affecting  the production and/or decay rates in a relevant way. Additionally, the presence of a SM-like Higgs boson in the allowed mass range is consistent with constraints coming from precision electroweak  data~\cite{ALEPH:2010aa}--\cite{LopesdeSa:2012ak}, and  therefore extensions of the SM which induce weak effects on precision electroweak observables are favored. The ATLAS experiment has further constrained a SM-Higgs boson with mass below 122.5~GeV, apart from a narrow window around 118~GeV. Moreover, an interesting excess of events, consistent with the production of a Higgs boson with a mass of about 125~GeV has been reported by both CMS and ATLAS. Most of the significance comes from the diphoton production rate, which, in spite being consistent with the SM prediction at the 2-$\sigma$ level, is  somewhat larger than expected from a SM Higgs boson. The Tevatron experiments also see an excess of events, consistent  with the production of a (115-135) GeV SM-like Higgs boson in associated production with vector gauge bosons, decaying into bottom quarks~\cite{TEVNPH:2012ab}.

The range of masses at which the Higgs-like signatures are observed is also consistent with the Higgs mass range predicted in the minimal supersymmetric extension of the Standard Model~(MSSM)~\cite{reviews}, with third generation squarks at the TeV scale and a large mixing parameter, $A_t$.  Indeed, due to the relation of the tree-level Higgs quartic coupling with the weak gauge couplings, the MSSM predicts a relatively light SM-like Higgs boson~\cite{Higgs:1964pi},\cite{Higgs:1966ev} with a mass of the order of the weak gauge boson masses. The precise value of this Higgs mass is strongly dependent on loop corrections which depend quartically on the top quark mass and logarithmically on the scale of the stop masses. The SM-like Higgs mass has also relevant quadratic and quartic dependences on the stop mixing parameter $A_t$. For both stop masses at the TeV scale, there is a maximal value for the SM-like Higgs mass, which has been computed at the one and two-loop level by different methods, and is about 130~GeV~\cite{Okada:1990vk}--\cite{Degrassi:2002fi}.

There have been many articles interpreting the Higgs mass range of about 125~GeV in minimal supersymmetric extensions of the SM~\cite{Arbey:2011ab}--\cite{Brummer:2012ns}. Regarding the rates, it is worth noticing that although the best fit value of the diphoton production rate is larger than the one expected for a SM Higgs with a mass close to 125~GeV, the uncertainties are still large and the enhancement of the rate with respect to the SM expectation is only slightly more than a standard deviation at each experiment. In this article, following Ref.~\cite{Carena:2011aa}, we shall analyze the possibility that the observed diphoton rate enhancements are not a statistical fluctuation, but a result of the presence of light supersymmetric particles. In Ref.~\cite{Carena:2011aa} it was observed that in minimal supersymmetric models,  a large Higgs diphoton decay rate may be obtained in the presence of light staus, with large left-right mixing. Such large mixing is obtained for large values of the ratio of vacuum expectation values, $\tan\beta$, and a moderate to large Higgsino mass parameter, $\mu$.  Values  of the Higgs diphoton decay rate as large as $\sim$1.5--2 times the expected SM Higgs decay rate may be obtained for stau masses close to the LEP limit~\cite{LEPlimit}, without affecting other Higgs production modes in any significant way.     

In this article, we shall analyze the phenomenological properties of this scenario, without assuming any particular high energy supersymmetry-breaking structure.  In section 2, we review the relevance of light staus and summarize their main properties, including the current experimental limits, their impact on the Higgs rate and on precision electroweak observables. We also discuss the corrections to the muon anomalous magnetic moment in the presence of light sleptons. In section 3 we concentrate on the predictions for the Dark Matter relic density. In section 4 we study the renormalization group evolution of the supersymmetry-breaking parameters, and study the constraints on the messenger scale obtained from demanding flavor universality of the scalar mass parameters at this scale. In section 5 we concentrate on possible searches at the LHC of light staus, with sizable left- and right-handed components.  We focus on searches for weakly associated production of staus and sneutrinos and weakly produced pairs of staus, since these are the channels that would probe this scenario independently of the mass of other supersymmetric particles. We reserve section 6 for our conclusion.

\section{\boldmath{Sfermion Effects on the~$h\to\gamma\gamma$~Rate}}

The three main effects on the $h\to \gamma\gamma$ rate are
\begin{itemize}
\item{Squark  effects on the gluon fusion rate  and on the $\gamma\gamma$ Higgs width.}
\item{Light stau effects on the Higgs diphoton decay width due to stau mixing effects controlled by $\mu \tan \beta$.}
\item{Higgs mixing effects controlled by $A_{\tau}$ and $m_A$, modifying the Higgs $b\bar b$ decay width leading to a suppression/enhancement of both the  $\gamma\gamma$ and $ZZ$ Higgs rates.  }
\end{itemize}

In this section we shall expand on these three different effects.

An enhancement of the branching ratio of the Higgs decay into photons can be obtained through the presence of light third generation stops in the presence of large mixing. However, in general such a modification is overcompensated for by a decrease of the gluon fusion production rate~\cite{Spira:1995rr}--\cite{Grazzini}. Hence for stops at the TeV scale, the production of photons in association with the Higgs tends to be reduced. One could also consider the possibility of increasing the gluon fusion production cross section. However, such an enhancement, in the presence of a very light stop with small mixing,  demands that the heaviest stop mass be very large to achieve  a Higgs mass in the 125~GeV range. Additionally, this situation also tends to lead to a suppression of the branching ratio of the Higgs decay into two photons~\cite{Djouadi:1998az},\cite{Dermisek:2007fi}. Therefore, the $W^+W^-$ and $ZZ$ production rates tend to be enhanced more than the $\gamma \gamma$, in some tension with current LHC results~\footnote{Similar arguments can also be applied to third generation down-type squarks. Light sbottoms with small mixing can enhance the diphoton rate, due to the enhancement of the gluon fusion production rate, however the $W^+W^-$ and $ZZ$ channels would be even more enhanced than the $\gamma\gamma$ channel.}.

\begin{figure}
\begin{center}
\begin{tabular}{c c}
\includegraphics[width=0.44\textwidth]{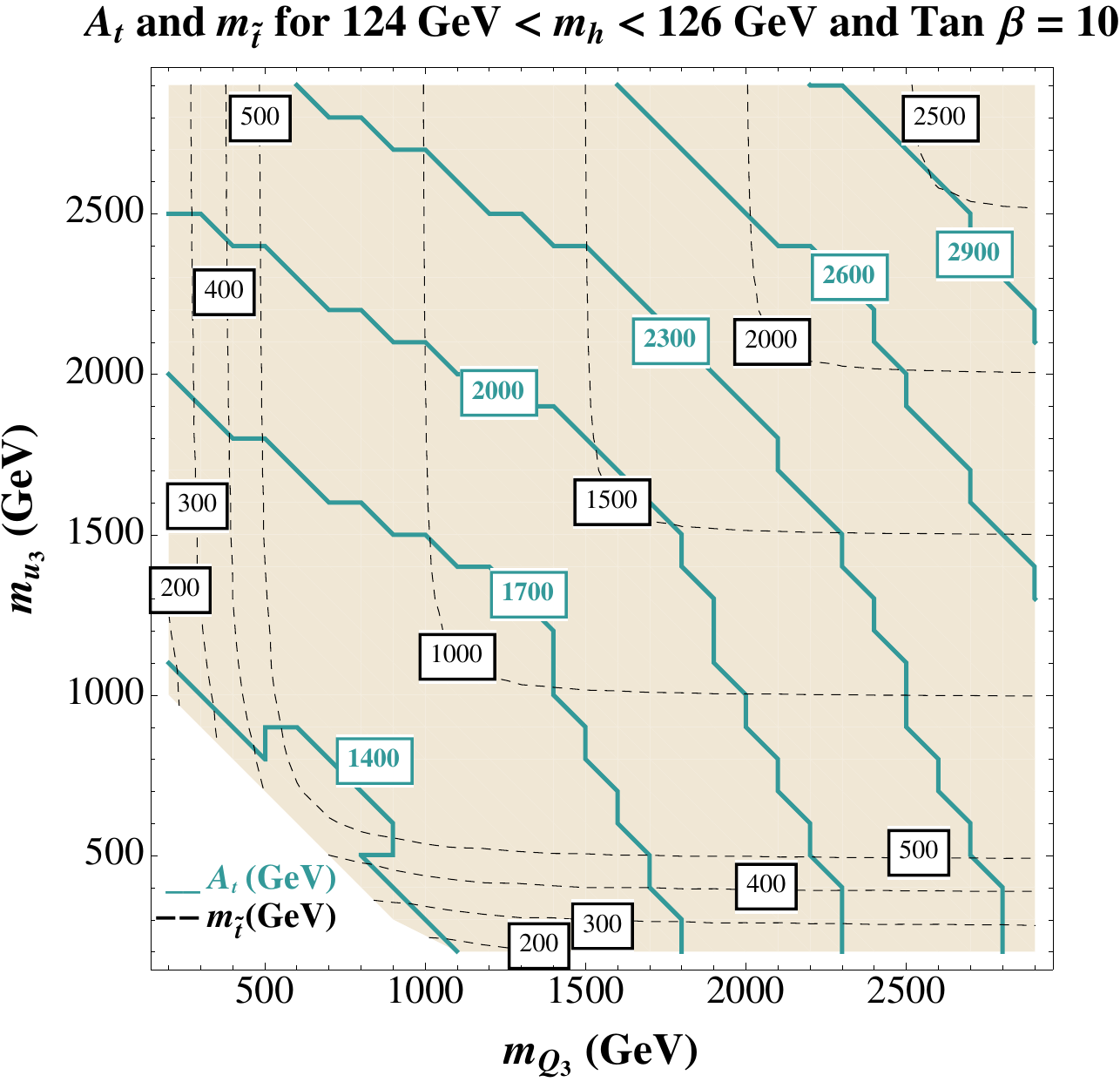}  &
\includegraphics[width=0.45\textwidth]{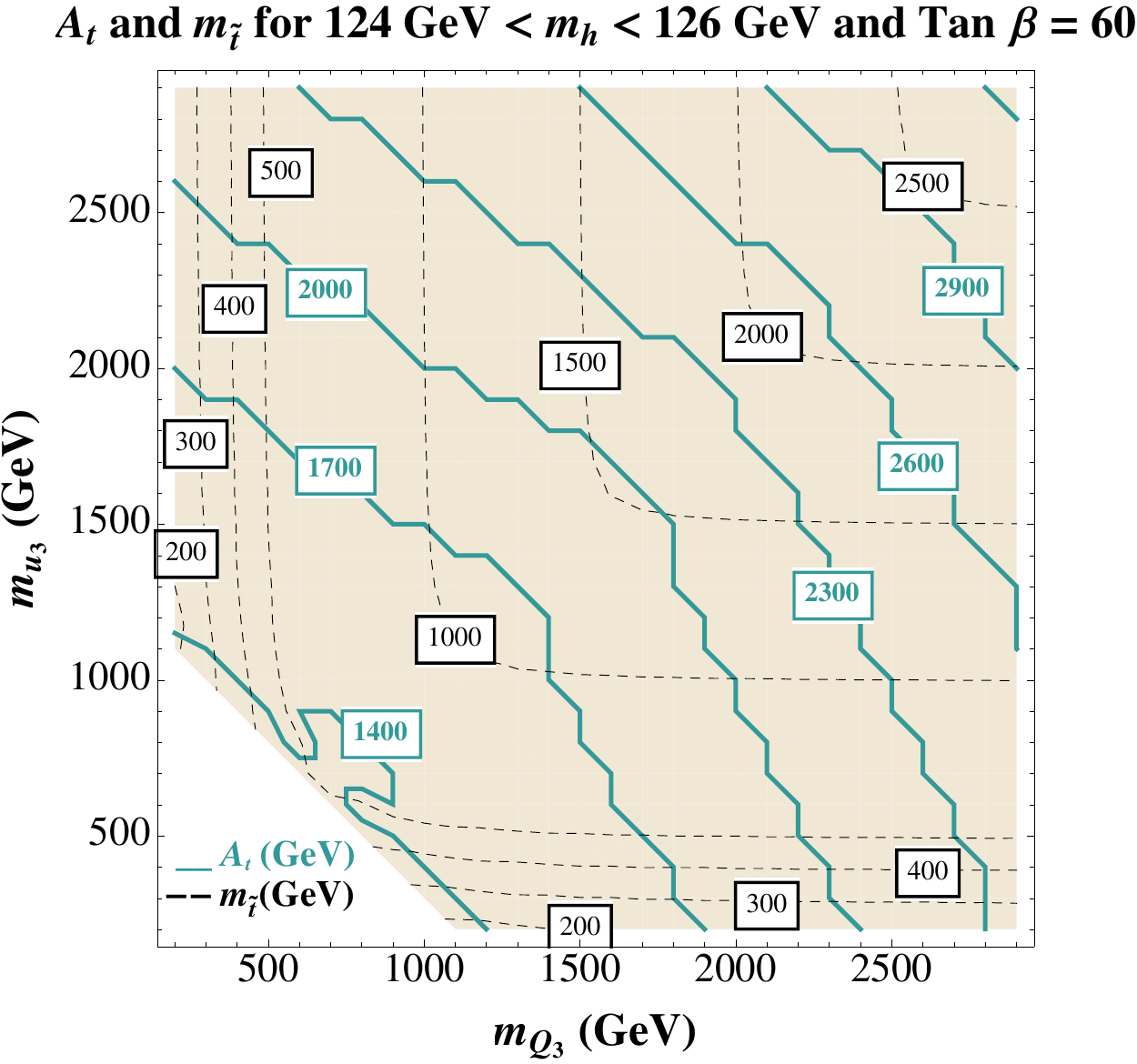}  \\
\end{tabular}
\end{center}
\caption{Contours of the stop mixing parameter, $A_t$, necessary for a Higgs mass $\sim$ 125~GeV given in the plane of the left- and right-handed stop soft supersymmetry-breaking mass parameters, $m_{Q_3}$, $m_{u_3}$ for $\mu = 650$~GeV, $m_A=1500$ GeV and $A_\tau=500$ GeV. \textit{Left:} $\tan\beta = 10$. \textit{Right:} $\tan\beta = 60$, which is where stau effects can be relevant for the diphoton production rate.}
\label{Higgsstop}
\end{figure}

On the other hand, relatively light third generation stops, with large mixing parameter, $A_t$, are consistent with a  $m_h \simeq 125$~GeV Higgs mass value.  The stop effects on the Higgs mass have been calculated by many authors in the past~\cite{Arbey:2011ab}--\cite{Brummer:2012ns}.  For  completeness, we have computed the Higgs mass constraints on the stop mass parameters  with \texttt{FeynHiggs}~\cite{FH},\cite{Frank:2006yh}, which includes the negative light stau effects. We have also verified the consistency of these results with a modified version of \texttt{CPsuperH}~\cite{CPsuperH}, which includes the same effects. The on-shell scheme values of the stop parameters necessary to achieve a Higgs mass in the range (124--126)~GeV are shown  in Fig.~\ref{Higgsstop}, for $\tan\beta = 10$ (left panel) and for $\tan\beta = 60$ (right panel), with  $\mu = 650$~GeV, $m_{L_3} \simeq m_{e_3} \simeq 280$~GeV and $A_\tau=500$ GeV.  The comparison of the results for $\tan\beta = 10$ and $\tan\beta = 60$ shows that the slight gain in the Higgs mass that is obtained by the increase in $\tan\beta$ is compensated for by the small negative effects associated with light staus~\cite{Carena:2011aa} with relatively large values of $\mu$.  Both stops can get masses smaller than 1~TeV. On the other hand, one of the stops can acquire a mass as small as  about (100--200)~GeV, provided that the heaviest stop mass and the stop mixing parameter are somewhat larger than 1~TeV, $A_t \simeq m_{Q_3} \simgt  1.5$ TeV. 

A positive contribution to the $\gamma\gamma$ production rate, without modifying the gluon fusion rate, may only be due to loops of sleptons and charginos. Due to the fact that charginos couple with $\tan\beta$ suppressed weak coupling strength to the Higgs, their impact on the branching fraction of the Higgs decay to photons is very small, of the order of a few percent at values for $\tan\beta \geq 5$~\cite{Diaz:2004qt}.  Similarly, first and second generation sleptons do not induce a relevant change of this branching ratio.However,  at large values of $\tan\beta$, staus have an enhanced coupling to the SM-like Higgs, induced by the Higgsino mass parameter, $\mu$.
Large values of $\mu$ and $\tan \beta$  induce large mixing in the stau sector leading to an enhancement of the branching ratio of the Higgs decay into photons. This behavior can be understood by looking at the stau mass matrix, given by
\begin{eqnarray}
\mathcal M_{\tilde{\tau}}^2
\simeq
\left[
\begin{tabular}{c c}
$ m_{L_3}^2 + m_{\tau}^2 + D_L$  &
$ h_{\tau}v(A_\tau\cos\beta - \mu \sin\beta) $ \\
$ h_{\tau}v(A_\tau\cos\beta- \mu \sin\beta)$ &
$ m_{E_3}^2 + m_{\tau}^2 + D_R $\\
\end{tabular}
\right]\;,
\end{eqnarray}
where $h_\tau\simeq m_\tau/(v\cos\beta)$ and $D_L$ and $D_R$ are the D-term contributions to the slepton masses~\cite{reviews}. At large values of $\tan\beta$ these are approximately given by $D_i \simeq -(T_3^i - Q^i \sin^2\theta_W) m_Z^2$, where $T_3^i$ and $Q^i$ are the $SU(2)$ isospin component and electromagnetic charge of the corresponding taus. 

However, for the mixing effects to be relevant, another condition  must be fulfilled: the lightest stau has to be rather light, with a mass close to the LEP limit. 
 
An intuitive way of understanding the effects of light staus with large mixing  on the $h\to\gamma\gamma$ rate is related to the fact that the contribution to the diphoton amplitude including  scalar particles with masses comparable or larger than the Higgs mass is approximately proportional to~\cite{Ellis:1975ap},\cite{Shifman:1979eb}
\begin{equation}
A_{\gamma\gamma}^{\textrm{SM}}+\Delta A_{\gamma \gamma} \propto -13 + \frac{Q_S^2}{3}\frac{\partial{\log\left[\det M_S^2(v)\right]}}{\partial\log(v)}\;,
\end{equation}
where $-13$ comes from the SM contribution. Here $Q_S$ are the charges of the scalar particles and $M_S^2$ is the mass matrix.  A relevant  contribution to $\Delta A_{\gamma \gamma}$ must be negative and of order one to contribute in a significantly constructive way to the dominant $W^{\pm}$ amplitude. For staus, ignoring subleading terms,  the above expression is approximately given by
\begin{equation}
\Delta A_{\gamma \gamma} \propto - \frac{2 (\mu \tan\beta)^2 m_{\tau}^2}{3\left[ m_{L_3}^2 m_{e_3}^2 - m_{\tau}^2 (\mu \tan\beta)^2\right]}.
\label{gammagamma}
\end{equation} 
For equal soft breaking masses for the left and right-handed staus, this simplifies to
\begin{equation}
\Delta A_{\gamma \gamma} \propto - \frac{m_{\tilde{\tau}_2^2}}{6 \; m_{\tilde{\tau}_1^2}} \left( 1 - \frac{m_{\tilde{\tau}^2_1}}{m_{\tilde{\tau}_2^2}} \right)^2.
\end{equation}
In order to get a sizable contribution, the splitting of the stau masses should then be such that $m_{\tilde{\tau}_2}/m_{\tilde{\tau}_1} \simgt 3$. Since we assume $m_{L_3}=m_{e_3}$ and $\tan\beta = 60$, for a light stau mass of order 100~GeV this can only be obtained for $\mu \simgt 300$~GeV (or,  in general, $\mu \simgt 300\; {\rm GeV} (60/\tan\beta)$). Larger values of $\mu\tan\beta$ lead to stronger effects, and for values of $\mu \simeq 1$~TeV and $\tan\beta \simeq 60$, enhancements of the rate of order 2 may be obtained.  In this article, we will work with more moderate values of $\mu \simeq 650$~GeV, and $\tan\beta \simeq 60$, for which enhancements of order 50\% may be obtained (see Fig. 2)~\footnote{In Ref.~\cite{Carena:2011aa}, we noticed that there was a small discrepancy in the computation of the Higgs diphoton rate between the results obtained by \texttt{FeynHiggs} and \texttt{CPsuperH}~\cite{CPsuperH}. This has been resolved and we display the corrected results.}. Note that in our analysis we are always taking $m_A$ beyond the current exclusion bound coming from $A,H\to\tau\bar\tau$ searches~\cite{Aad:2011rv}--\cite{CMStau}\footnote{ We recently noted that the bound on $m_A$ may be modified analyzing the $A,H\to b\bar b$ channel with a Higgs produced in association with a $b$ quark at the LHC~\cite{Carena:2012rw}.}. 

\begin{figure}
\begin{center}
\begin{tabular}{c c}
(a) & (b)\\
&\\
\includegraphics[width=0.45\textwidth]{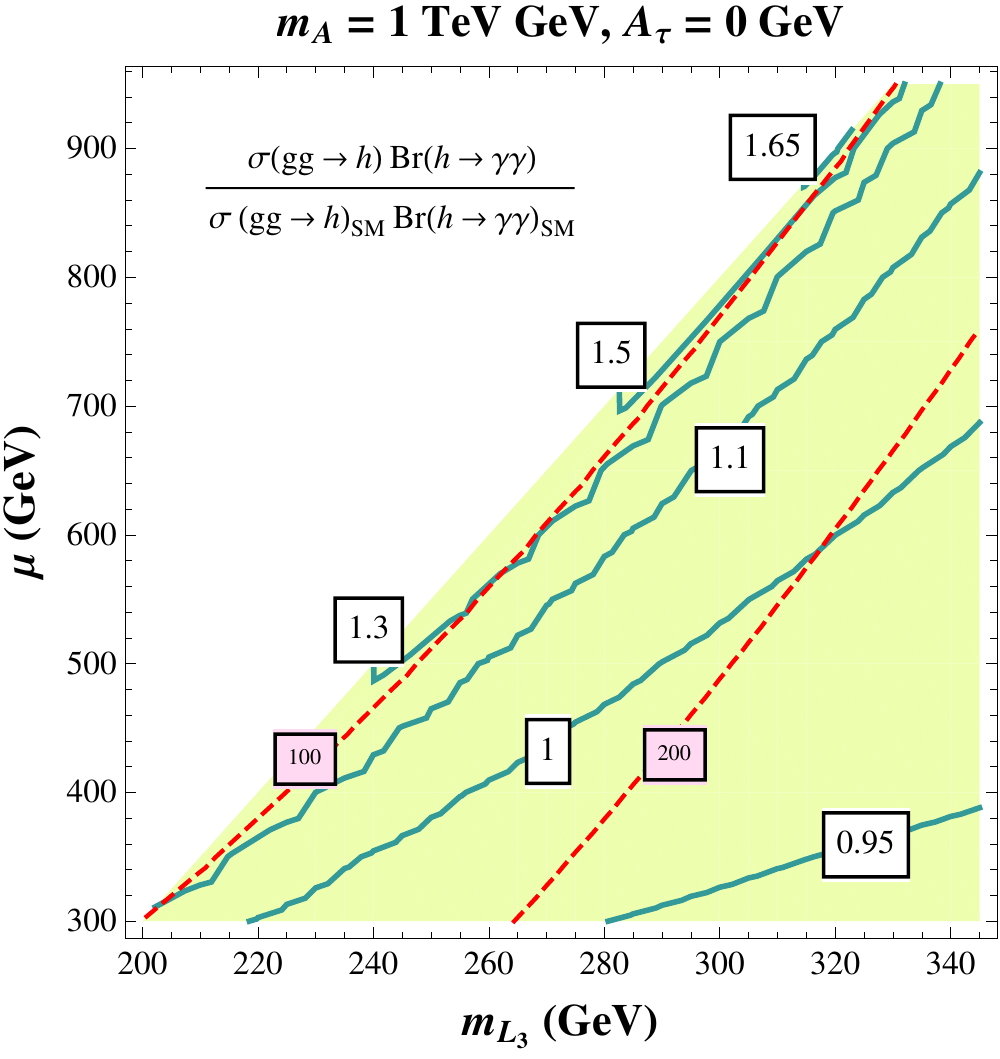}  &
\includegraphics[width=0.45\textwidth]{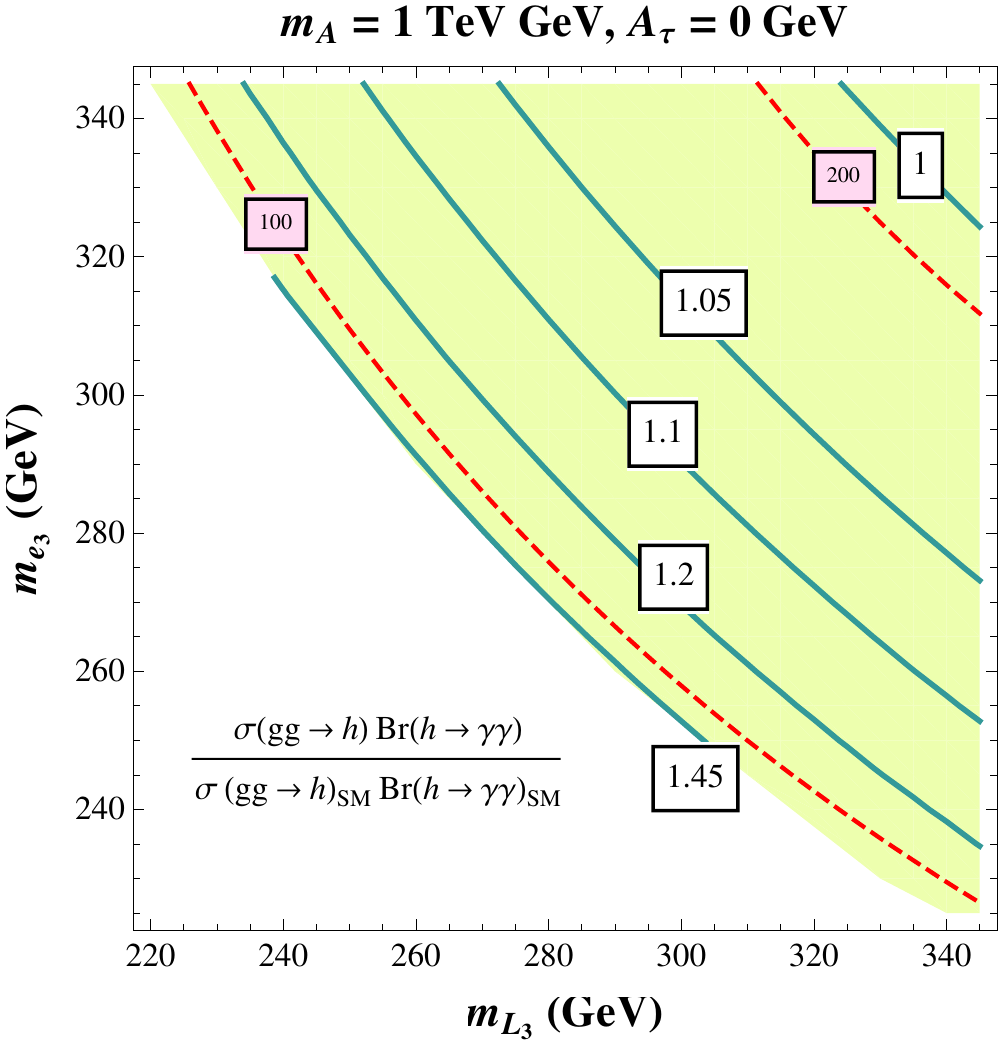}  \\
&\\
(c) & (d)\\
&\\
\includegraphics[width=0.45\textwidth]{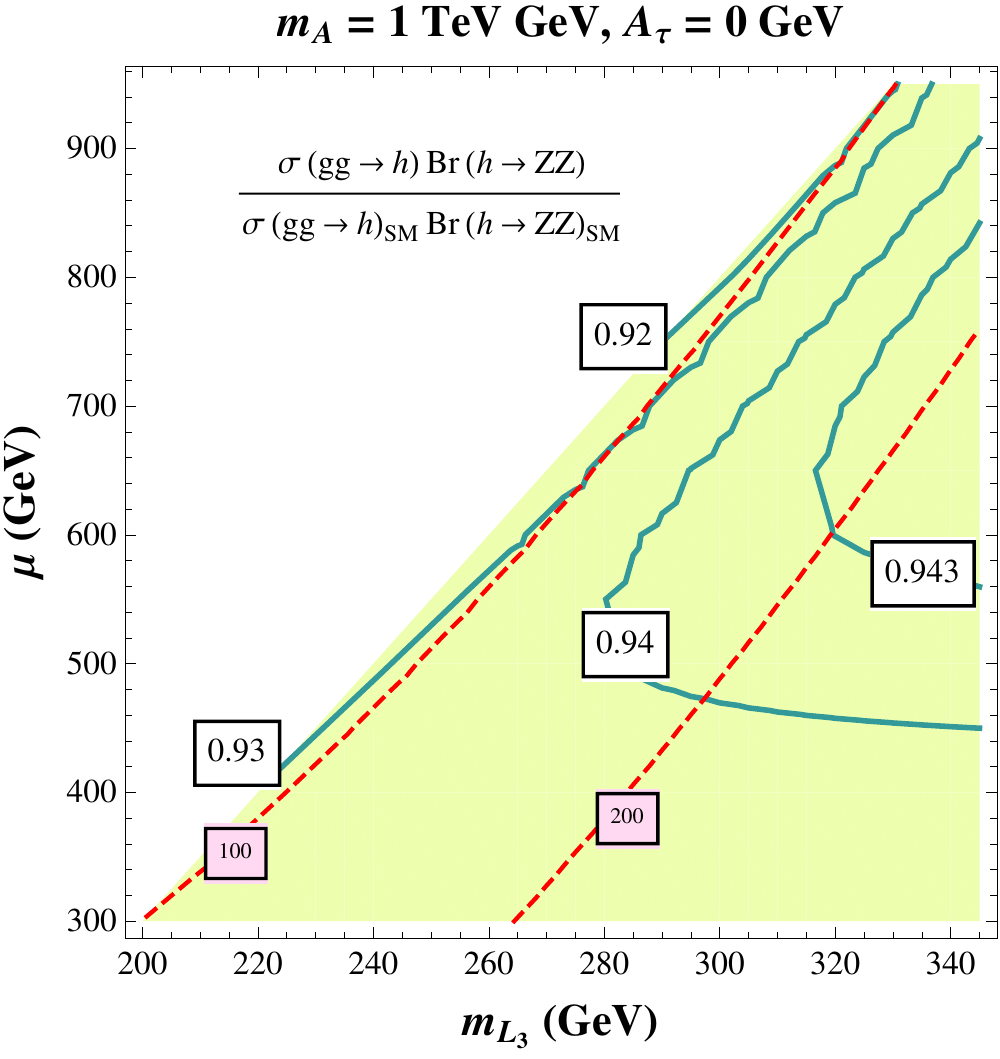}  &
\includegraphics[width=0.45\textwidth]{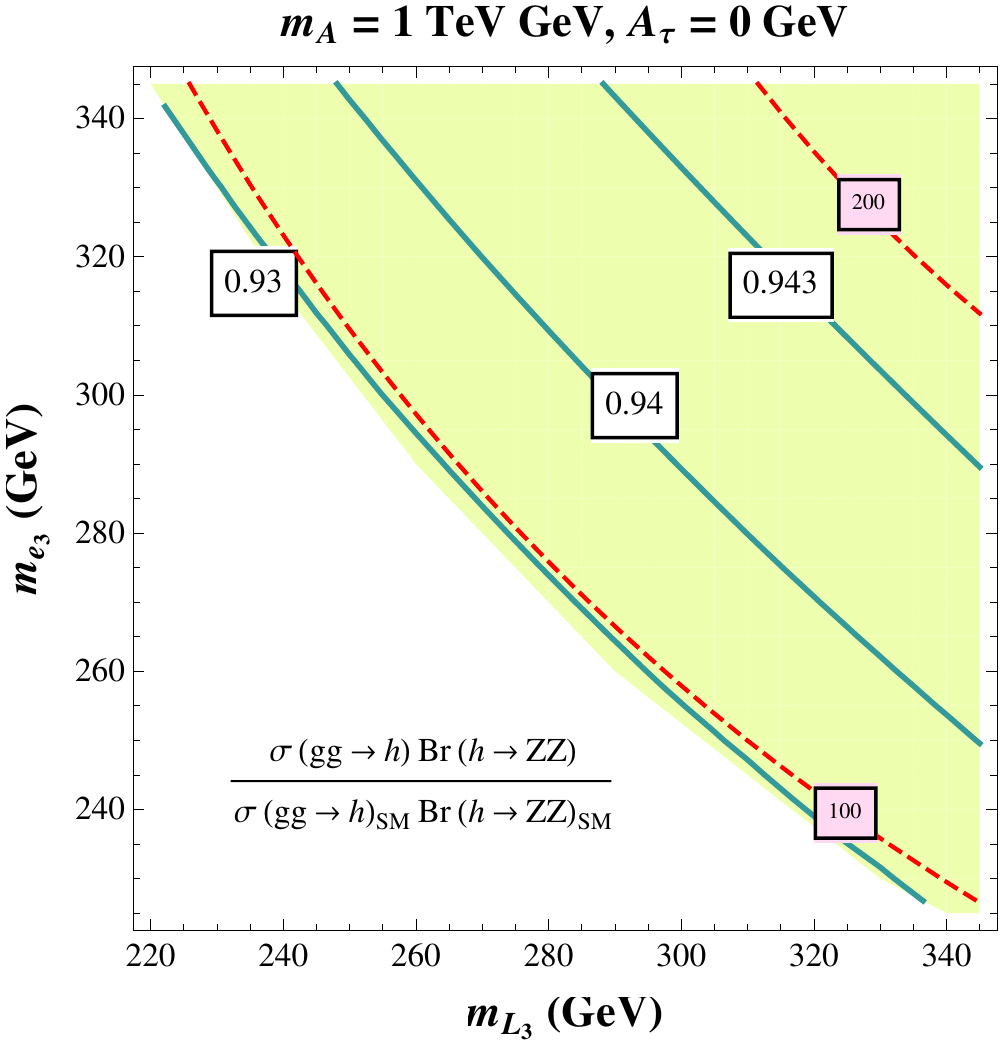}\\  
\end{tabular}
\end{center}
\caption{Contour plots of the ratio of  the $\sigma(gg \to h) \times $ BR($h \to VV$) to its SM value, in the \textit{(a) \& (c):} $\mu$--$m_{L_3}$ plane with $m_{e_3}=m_{L_3}$, and \textit{(b) \& (d):} $m_{e_3}$--$m_{L_3}$ plane with $\mu =650$ GeV. $\tan\beta = 60$, $m_A=1$ TeV and $A_\tau=0$ GeV are kept fixed for all the plots.  The relevant squark parameters are $A_t=1.4$ TeV and $m_{Q_3}=m_{u_3}=850$ GeV giving $m_h\sim 125$ GeV. Red dashed lines are contours of lightest stau masses. The yellow shaded area denotes the region satisfying the LEP bound on the lightest stau mass. Enhanced branching ratios are obtained for values of $\mu$ for which the lightest stau mass is close to its experimental limit, of about (85-90)~GeV. }
\label{gammastau}
\end{figure}

The dependence of $\sigma(gg \to h) \times BR(h\to \gamma\gamma)$ in the $m_{L_3}$--$m_{e_3}$ plane, for $\tan\beta = 60$, $\mu = 650$~GeV, $A_\tau = 0$~GeV, $m_A=1$ TeV , as well as in the  $m_{L_3}$--$\mu$ plane for $m_{L_3} = m_{e_3}$ is shown in Fig.~\ref{gammastau}. Solid lines represent contours of the diphoton rate, normalized to the SM value. Dashed lines represent contours of the lightest stau mass. The squark sector is fixed at $m_{Q_3} = m_{u_3} = 850$~GeV and $A_t = 1.4$~TeV,  consistent with a Higgs mass of about 125~GeV. 
 A clear enhancement  of the order of $50\%$ in the total photon rate production  is observed in the region of parameters leading to light staus, close to the LEP limit.   
 For  this set of parameters the Higgs mixing effects, as well as the effects coming from the stop sector, are small. Hence, both the production rate of weak gauge bosons, shown in Fig.~\ref{gammastau}, and the branching ratio of the Higgs decay into bottom quarks, remain very close to the SM ones. 

\begin{figure}
\begin{center}
\includegraphics[width=0.4\textwidth]{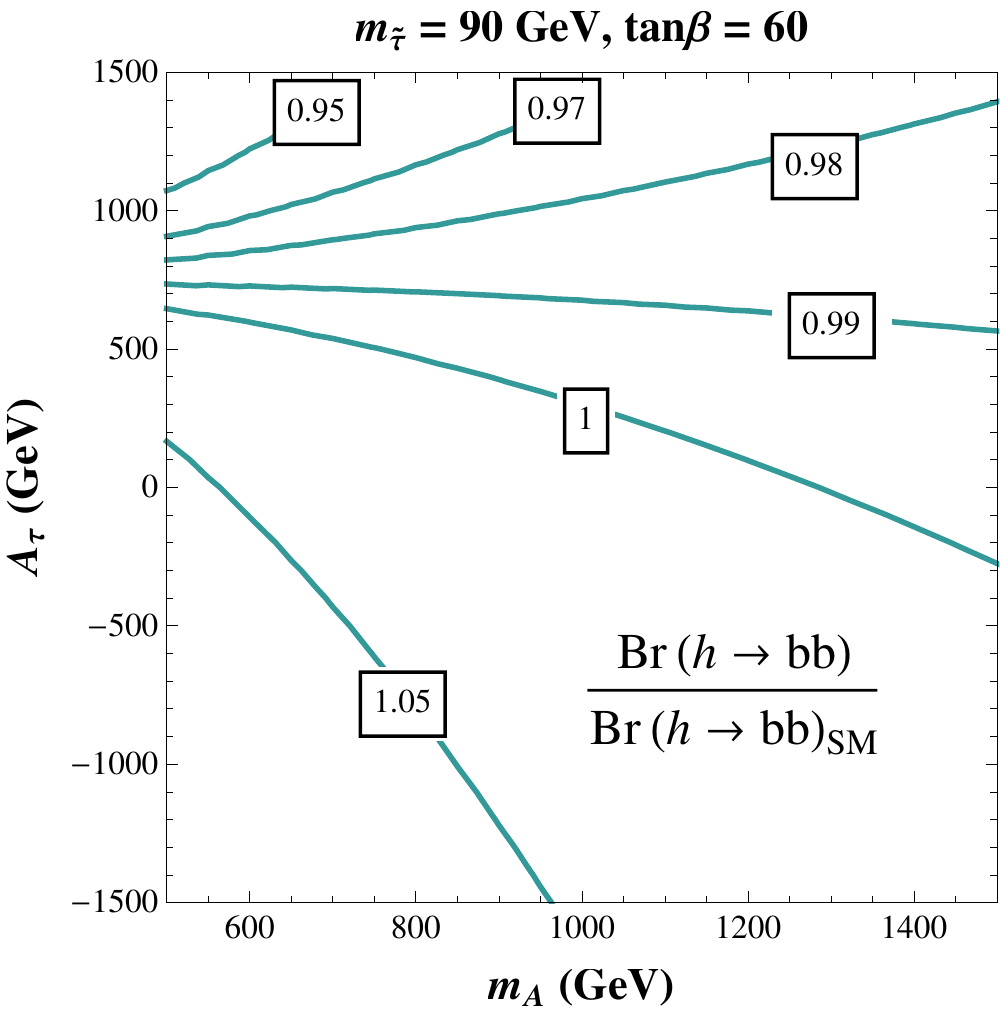}
\caption{\label{h}
Contour plots of the ratio of BR($h \to b\bar b$) to its SM value, in the $m_A$--$A_\tau$ plane with $\tan\beta = 60$, $m_{e_3}=m_{L_3}=250$ GeV. We fix $m_{\tilde{\tau}_1}=90 $ GeV, hence $\mu$ varies in the range 500--550~GeV. The relevant squark parameters are $A_t=1.8$ TeV and $m_{Q_3}=m_{u_3}=1.5$ TeV corresponding to $m_{\tilde{t}_{1,2}}\sim 1.4, 1.6$ TeV and $m_{h}\sim$ 125 GeV.  }
\end{center}
\end{figure}

As discussed briefly in Ref.~\cite{Carena:2011aa}, the mixing parameter $A_\tau$ and $m_A$ play a relevant role 
in the Higgs mixing and therefore in the width of the Higgs decay into bottom quarks. 
In particular, the mixing in the Higgs mass matrix is given by  $\sim-m_A^2 \sin \beta \cos \beta + Loop_{12}$, where $Loop_{12}$ includes the dependence on $A_\tau$~\cite{Carena:2011aa}. Large negative (positive) values of $A_\tau$ and moderate values of $m_A$ can lead to an enhancement (suppression) of the width of the Higgs decay into bottom quarks, and a subsequent suppression (enhancement) of the photon and weak gauge boson production rates. Contours of the  $BR (h\rightarrow b\bar b)$ normalized to the SM, are presented in Fig.~\ref{h} in the $m_A$-$A_\tau$ plane. The squark masses are all heavy so that they have a minimal impact, and therefore the effects shown in the plot are dominantly due to the Higgs mixing effects. We fix $m_{\tilde{\tau}_1}=90 $ GeV, with  $\tan\beta = 60$ and $m_{e_3}=m_{L_3}=250$ GeV and  hence $\mu$ varies in the range 500--550~GeV. As can be seen, smaller~(larger) values of $m_A$ allow for a larger~(smaller) variation of the $h\to b\bar b$ branching ratio due to $A_\tau$. Since the $b\bar{b}$ decay width is the dominant one for a Higgs with a mass of 125 GeV, the variation of $BR (h\rightarrow b\bar b)$  is relatively small. We verified that a 5\% change in $BR (h\rightarrow b\bar b)$ corresponds to approximately a 20\% variation of $\Gamma(h\to b\bar{b})$ with respect to the SM quantities.

\begin{figure}
\begin{center}
\begin{tabular}{c }
(a)\\
\\
\includegraphics[width=0.9\textwidth]{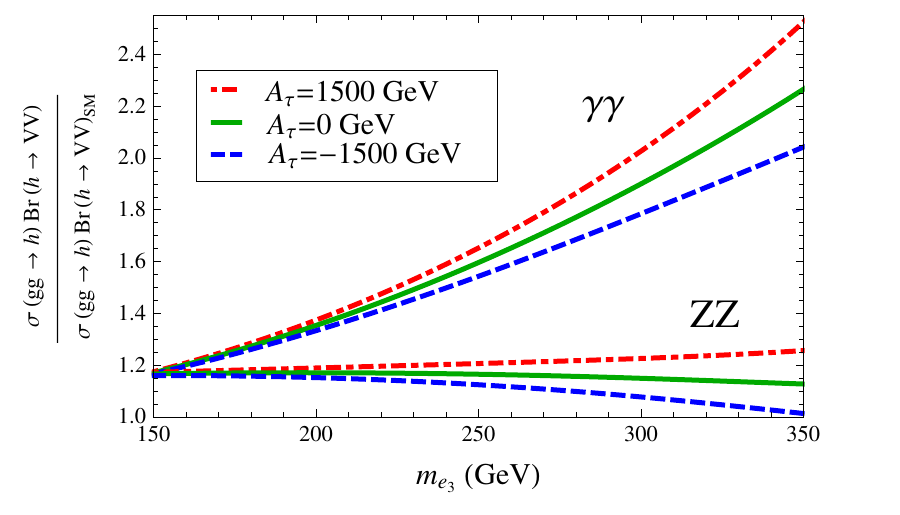}  \\
\\
(b)\\
\\
\includegraphics[width=0.9\textwidth]{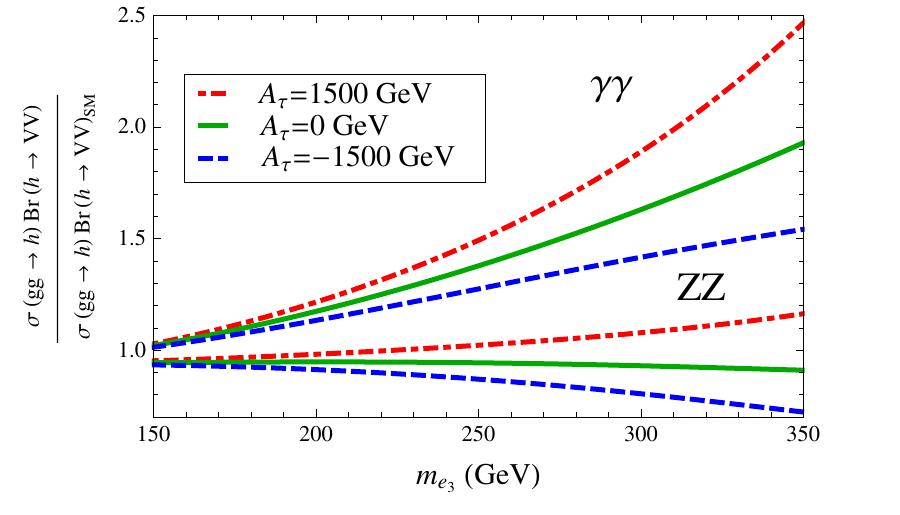}  \\  
\end{tabular}
\end{center}
\caption{ Ratio of  the $\sigma(gg \to h) \times $ BR($h \to VV$) to its SM value, for both $V=\gamma$ and $V=Z$ as a function of $m_{e_3}=m_{L_3}$, for $\tan \beta=60$ varying $\mu$ such that  $m_{\tilde{\tau}_1}=90$ GeV for different values of $A_\tau$. The Higgs mass varies with $m_{e_3}$, but remains $\sim125$ GeV. \textit{(a)}: $m_A = 1.5$ TeV,  $A_t = 2$ TeV, $m_{Q_3} =2.5$ TeV, $m_{u_3} = 100$ GeV leading to $m_{\tilde{t}_1}\sim140$ GeV.  \textit{(b)}: $m_A =1$ TeV, $A_t = 1.4$ TeV, $m_{Q_3} =1.5$ TeV, $m_{u_3} = 500$ GeV leading to $m_{\tilde{t}_1}\sim500$ GeV.}
\label{BrRat}
\end{figure}



Fig.~\ref{BrRat} summarizes the above discussed effects on the diphoton rates. In this figure we present  the diphoton and weak vector boson production rates as a function of the slepton masses, with $m_{L_3} = m_{e_3}$. Each plot shows the dependence on three different values of the stau mixing parameter $A_\tau$. We fix $\tan \beta=60$ and vary $\mu$ such that the lightest stau mass is 90~GeV. The CP-odd Higgs mass has been fixed to 1.5~TeV in Fig.~\ref{BrRat} (a) and to 1~TeV in Fig.~\ref{BrRat} (b). Moreover, in Fig.~\ref{BrRat} (a) we have chosen a case in which one stop is very light while the other is  heavy, while in Fig.~\ref{BrRat}(b), instead, the stops are  heavier. 
 
The effect of the sleptons on the diphoton rate depends on the ratio of $\mu/m_{e_3}$ (see Eq.~\ref{gammagamma}). To keep the stau mass constant at $90$ GeV, $\mu$ is very small for small values of $m_{e_3}$ ($\sim 150$ GeV) and hence the sleptons have no effect on the Higgs decay width into photons in this regime. Therefore, the values of the rates shown for the smallest values of $m_{e_3}$ in both the plots in Fig.~\ref{BrRat}  determined only by the squark sector effects. In Fig.~\ref{BrRat}(a), since the lightest stop is very light and  $A_t$ is smaller than the heaviest stop mass, $A_t<m_{Q_3}$, the gluon fusion rate is somewhat enhanced.  Instead in Fig.~\ref{BrRat}(b), the gluon fusion rate is close to the SM one, which is the generic expectation in the regime of squark parameters necessary to achieve a 125~GeV SM-like Higgs boson. Stops have the opposite effect on the diphoton width as on the gluon fusion rate. Therefore the light stop in Fig.~\ref{BrRat}(a) leads to a slight suppression of the Higgs diphoton width compared to Fig.~\ref{BrRat}(b). This is seen as the gap in Fig.~\ref{BrRat}(b) at $m_{e_3}\sim 150$ GeV, which is missing in Fig.~\ref{BrRat}(a), between the $\gamma\gamma$ and $ZZ$ rates.   

Since we keep the mass of the lightest stau and $\tan \beta$ fixed, as we increase $m_{e_3}$, $\mu/m_{e_3}$ is increasing for each constant value of $A_\tau$. Thus the enhancement of the $\gamma\gamma$ rate with increasing values of $\mu \tan \beta/m_{e_3}$ is illustrated in the figure. 

The parameter $A_\tau$ directly affects the CP-even Higgs mixing, and therefore the Higgs $b\bar b$ decay width. Positive values suppress this width and negative values enhance it.  The change in the $b\bar b$ width impacts both the $\gamma\gamma$ and $ZZ$ branching ratios in the same way, through a variation of the total Higgs width.  Additionally,  the CP-even Higgs mixing effects depend on $A_\tau \mu^3$~\cite{Carena:2011aa} and hence, larger values of $\mu$ lead to a larger enhancement/suppression of the $\gamma\gamma/ZZ$ Higgs rates. Finally, mixing effects are suppressed by $m_A$ (see, for example, Ref.~\cite{arXiv:1107.4354}). We note that if one has smaller values of $m_A$ for the same value of $\mu \tan \beta$, one should expect a larger difference for the rates between negative and positive values of $A_\tau$ . This can be seen by comparing the spread in the rates from $A_\tau=-1.5$ to 1.5 TeV, in both plots in Fig.~\ref{BrRat} for a given value of $m_{e_3}$. As can be clearly seen, the spread in (a), corresponding to $m_A=1.5$ TeV is much smaller than in (b) where $m_A=1$ TeV.

 One interesting observation is that for negative values of $A_{\tau}$ and light staus, one can obtain an enhancement of the bottom quark width, resulting in a  $ZZ$ and $WW$ weak boson production rate that is of the order or smaller than the SM values, while keeping an enhanced Higgs diphoton rate.  Such modified branching ratios would be in qualitative agreement with the signals observed by the Tevatron and LHC experiments~\cite{ATLAS:2012ae},~\cite{Chatrchyan:2012tx},~\cite{TEVNPH:2012ab}.

A word of caution is in order here. Large values of the stau mixing may lead to the presence of additional minima in the scalar potential, spoiling the stability of the electroweak vacuum.  In such a case, one should demand that the physical vacuum has a lifetime larger than the one of the Universe.  A first study of the metastability condition has been done in Ref.~\cite{Hisano:2010re} at the tree level, leading to an interesting constraint on $\mu\tan\beta$ which is satisfied by approximately
\begin{equation}
|\mu \tan\beta| \simlt 40 \left(\sqrt{m_{L_3}} + \sqrt{m_{E_3}}\right)^2 -  10^4 \; {\rm GeV}.
\end{equation}
This constraint would disallow the possibility of a light stau, with a mass close to the LEP limit for mass parameters $m_{e_3} \simeq m_{L_3} \simgt 250$~GeV, and therefore  enhancements of the diphoton rate due to staus larger than about 50\%.  The analysis of Fig.~\ref{BrRat} shows that  enhancements of the diphoton rate up to 80 \% may be obtained in region of parameters where the above bound is violated by $\lesssim 15\%$. Hence, it would be relevant to study the stability of the above bounds by analyzing the full, one-loop stau--Higgs effective potential.

\subsection{EWPTs and the Muon Anomalous Magnetic Moment}

\begin{figure}
\begin{center}
\begin{tabular}{c c}
\includegraphics[width=0.45\textwidth]{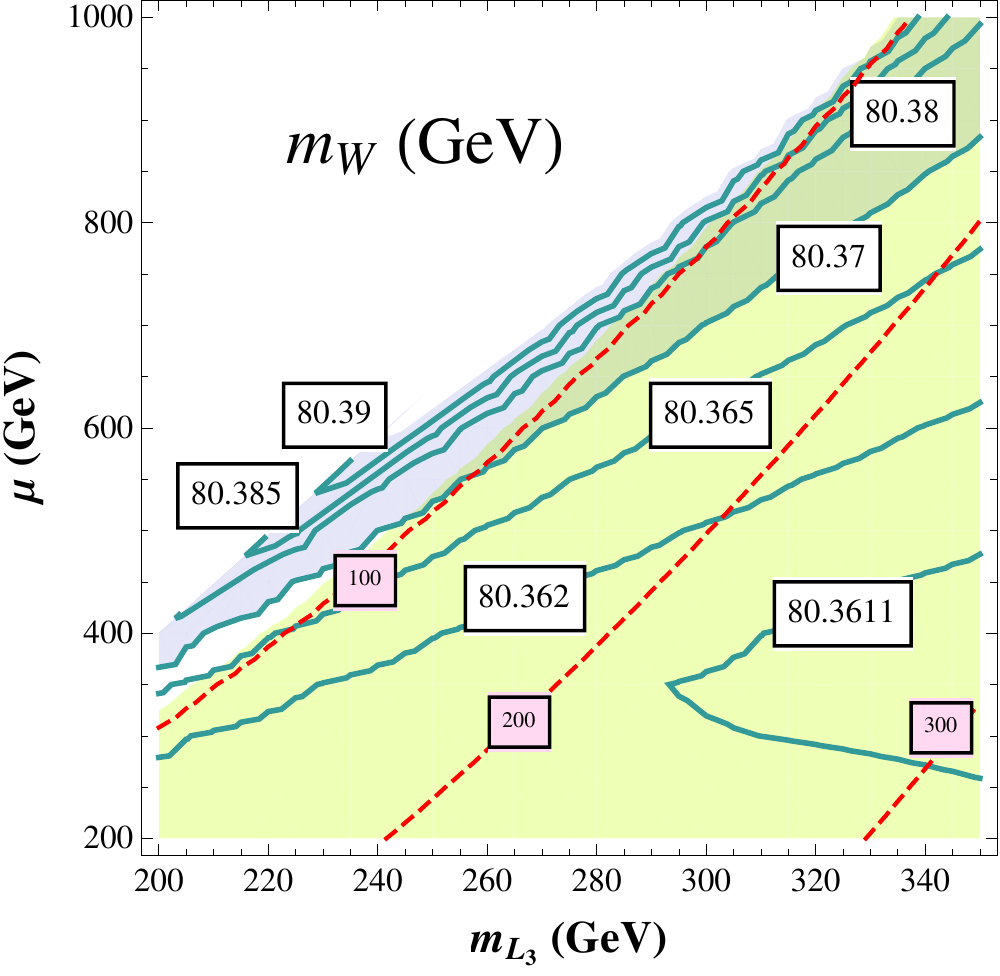}  &
\includegraphics[width=0.45\textwidth]{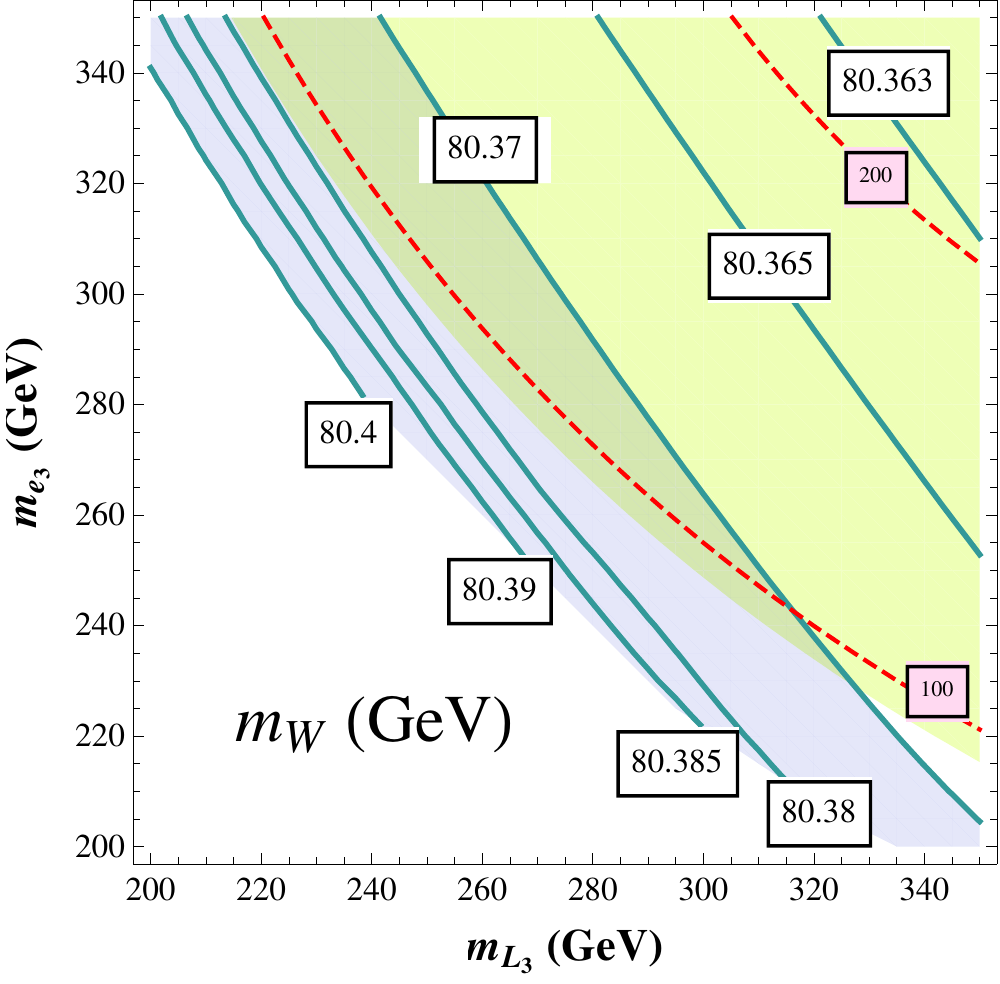}  \\
\end{tabular}
\end{center}
\caption{ Contour plots of $m_W$ Light blue fill denotes regions experimentally consistent within 1-$\sigma$ for the $W$ mass~($80.385\pm0.015$ GeV), with darker blue contours specifying the values of $m_W$. Light green fill denotes allowed region for the lightest stau mass~($m_{\tilde{\tau}} > 90$ GeV), with red lines denoting contours of the stau mass.  In the right panel, we present results for  $\mu = 650$~GeV and $\tan\beta = 60$, while in the left panel $\tan\beta = 60$ and $m_{L_3} = m_{e_3}$. All the other soft parameters are fixed to 2 TeV }
\label{MWstau}
\end{figure}

Light staus may have a relevant effect on precision electroweak measurements.  For instance, they modify the predicted value of $m_W$~\cite{Chankowski:1993eu},\cite{Heinemeyer:2006px} (or analogously of $\Delta\rho$), the central measured value of which is somewhat above the predicted value in the SM for a light Higgs with a mass of about 125~GeV~\cite{TevatronElectroweakWorkingGroup:2012gb},~\cite{LopesdeSa:2012ak}.  Fig.~\ref{MWstau} shows the predicted value of $m_W$ in the MSSM, in the region of light staus, for third generation squark values consistent with a SM-like 125~GeV Higgs. These values have been obtained from the prediction for $\Delta\rho$, which is related to the dominant contribution to $m_W$ by the relation~\cite{Heinemeyer:2004gx}

\beq
\Delta m_W\simeq\frac{m_W}{2}\frac{\cos^2\theta_W}{\cos^2\theta_W-\sin^2\theta_W}\Delta\rho\,,
\eeq
where $\theta_W$ is the weak mixing angle.

 In general, the squark sector gives a very small contribution to the $W$ mass in the region of parameters consistent with $m_h\simeq 125$~GeV. On the other hand, as shown in Fig.~\ref{MWstau}, light staus consistent with the enhanced Higgs diphoton decay rate, lead to an enhancement up to 40 MeV compared the SM value, $m_W^{SM} \simeq 80.36$ GeV.  The composition of these light staus is  important for the determination of the $m_W$ corrections. The larger the left-handed component (the smaller $m_{L_3}$), the larger the effect (see right panel of Fig.~\ref{MWstau}). Since the light stau effects on $m_W$ are positive, and of the order of the current experimental uncertainties, the  present measurement, $m_W \simeq 80.385 \pm 0.015$~GeV,  places only a marginal constraint on this scenario.  For instance, in the example of Fig.~\ref{MWstau}, models with $m_{L_3} \simeq m_{e_3} \simgt 350$~GeV and large values of $\mu$ may lead to values of $m_W$ larger than  80.40~GeV, which is 1~$\sigma$ above the current experimental bounds.  On the other hand, for values of $m_{e_3} \simeq m_{L_3} \simlt 350$~GeV, and a light stau mass close the LEP bound as required to enhance the Higgs diphoton width, one obtains values of $m_W$ that are in good agreement with current experimental constraints, and actually in better agreement than the SM predicted values.  The same is true for non-equal values of the slepton masses, for $\mu = 650$~GeV and $\tan\beta=60$, as shown in the right panel of Fig.~\ref{MWstau}.

Light sleptons may also affect the predicted value of the anomalous magnetic moment of the muon, $(g_{\mu}-2)$~\cite{Moroi:1995yh}--\cite{Feng:2001tr}. The anomalous magnetic moment is of interest since its current measured value differs by more than 3 standard deviations from the predicted value in the SM~\cite{Jegerlehner:2009ry},\cite{Nakamura:2010zzi} 
  \beq
 \Delta a_\mu= a_\mu^{\rm{exp}}-a_\mu^{\rm{SM}}\simeq(3\pm 1 )\times 10^{-9}\,.
 \eeq

%
%
\begin{figure}
\begin{center}
\begin{tabular}{c c}
\includegraphics[width=0.45\textwidth]{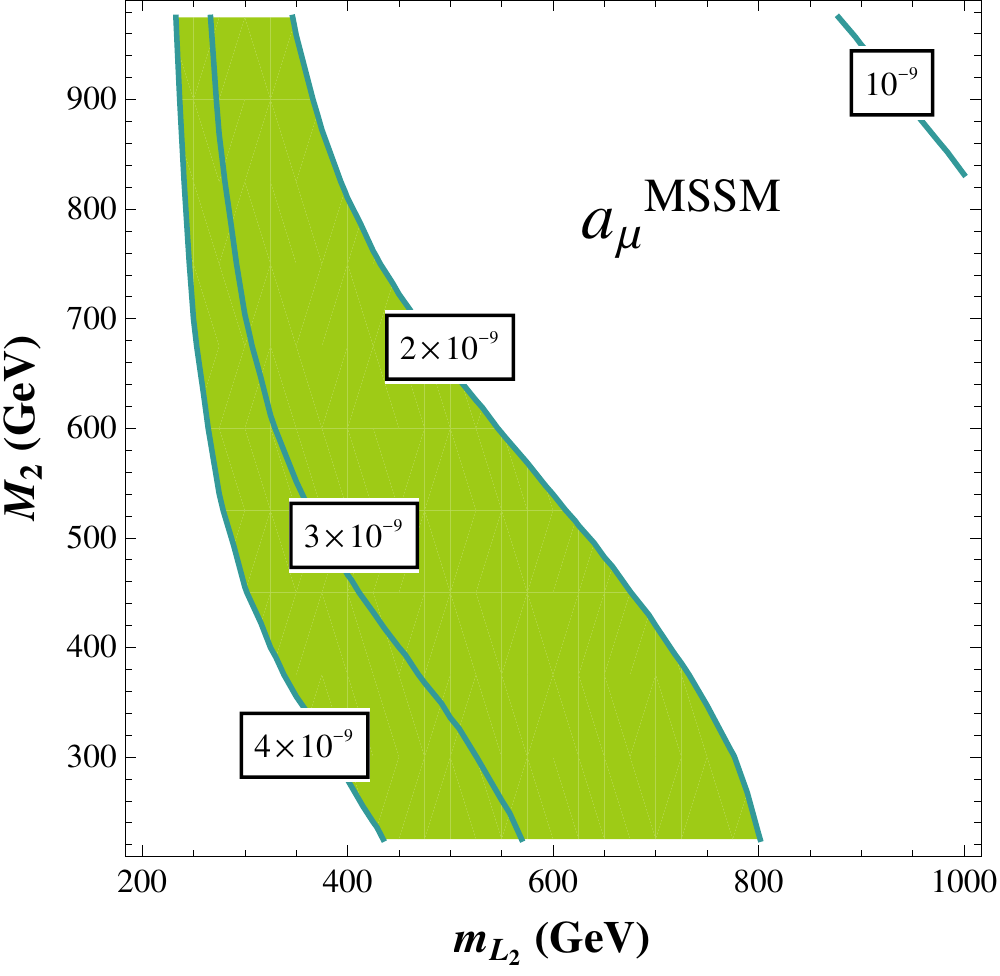}  &
\includegraphics[width=0.46\textwidth]{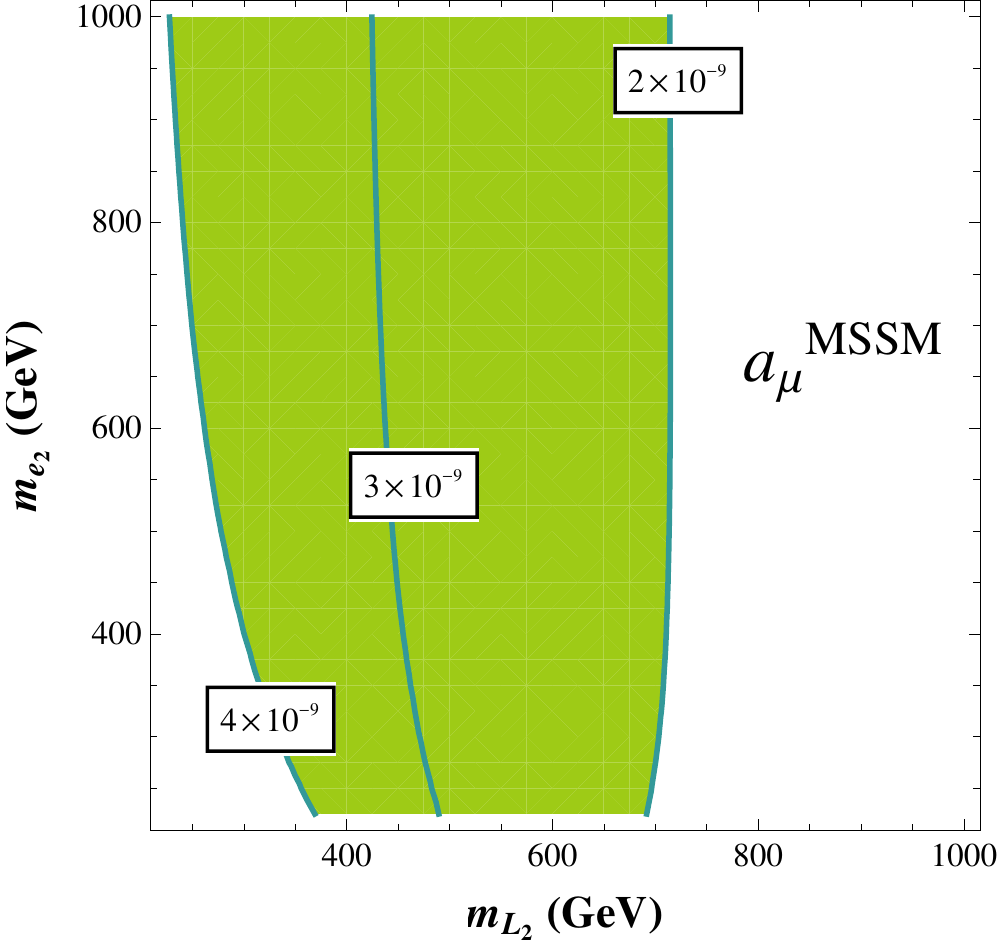}  \\
\end{tabular}
\end{center}
\caption{Contour plots of $a_{\mu}^{\rm MSSM}$, for $M_1$=35 GeV, $\mu = 650$~GeV and $\tan\beta = 60$. On the right panel, $M_2 = 400$~GeV, and on the left panel $\tan\beta = 60$ and $m_{L_2} = m_{e_2}$. Green fill denotes regions consistent with the experimental measurement within 1-$\sigma$. }
\label{g-2stau}
\end{figure}

  Although $(g_{\mu}-2)$ is not sensitive to the stau masses, it is interesting to investigate what would be the necessary value of the smuon masses in order to obtain a predicted value for $(g_{\mu}-2)$ consistent with the experimental value at the 1-$\sigma$ level.   The most important contribution to the anomalous magnetic moment comes from a diagram including charginos and muon sneutrinos.  Their effect is proportional to  $(M_2~\mu \tan\beta)$ and inversely proportional to the square of the slepton and chargino masses running in the loop. This contribution is given by the approximate expression~\cite{Carena:1996qa}
\begin{equation}
\frac{(g_{\mu}-2)^{\rm MSSM}}{ 2 \times 10^{-9}} =\frac{a^{\rm MSSM}_\mu}{ 1 \times 10^{-9}}  
\approx 1.5\left(\frac{\tan\beta }{10} \right) 
\left( \frac{300~\rm GeV}{m_{\tilde \ell}} \right)^2
{\rm{sign}}(\mu M_2)~,
\label{eq:g_2}
\end{equation}
valid under the hypothesis that the charginos and second generation sleptons are degenerate in mass. From Eq.~(\ref{eq:g_2}) we see that for $\tan\beta = 60$, charginos and sneutrinos with masses of about 500~GeV are necessary to get a value of $(g_\mu-2)$ close to its current measured central value. 

 Fig.~\ref{g-2stau} shows the predicted values of $a_{\mu}^{\rm{MSSM}}$ for different values of the smuon soft supersymmetry-breaking masses, for the values of $\mu$ and $\tan\beta$ for which light staus lead to an enhancement of the Higgs diphoton rate, for instance, for equal values of $m_{L_3} \simeq m_{e_3} \simeq 280$~GeV, $\tan\beta \simeq 60$ and $\mu \simeq 650$~GeV. We see from Fig.~\ref{g-2stau} that the values of the smuon soft supersymmetry-breaking parameters do not differ significantly from the analogous stau values. Since the chargino-sneutrino loop gives the dominant contribution to $(g_\mu-2)$, the result is mostly sensitive to the left-handed smuon mass parameter and only weakly sensitive to the right-handed smuon mass parameter, as shown in the right panel of Fig.~\ref{g-2stau}. For instance, for the values of $\mu = 650$~GeV and $\tan\beta=60$ represented in Fig.~\ref{g-2stau}, and for relatively small values of $M_2$, values of  $m_{L_2} \simeq  (500 \pm 200)$~GeV are necessary to obtain a value of $(g_{\mu}-2)$ within 1~$\sigma$ of  the experimental value. On the other hand, for large values of the chargino masses, there is still a subdominant contribution that is governed by smuon neutralino exchange proportional to $(M_1~\mu \tan\beta)$. 
This contribution is important only for small values of the left- and right-handed soft mass parameters and becomes relevant in determining the asymptotic value of the slepton soft mass parameters for large values of $M_2$ shown in the left panel of Fig.~\ref{g-2stau}.  

Observe that due to the small muon mass, the smuon mass eigenvalues are of the order of the soft supersymmetry-breaking parameters. Hence smuons, at least left-handed ones, tend to be heavier than the lightest stau.

\section{Dark Matter}\label{sec:DM}

\begin{figure}
\begin{center}
\begin{tabular}{c c}
(a) & (b) \\
\includegraphics[width=0.45\textwidth]{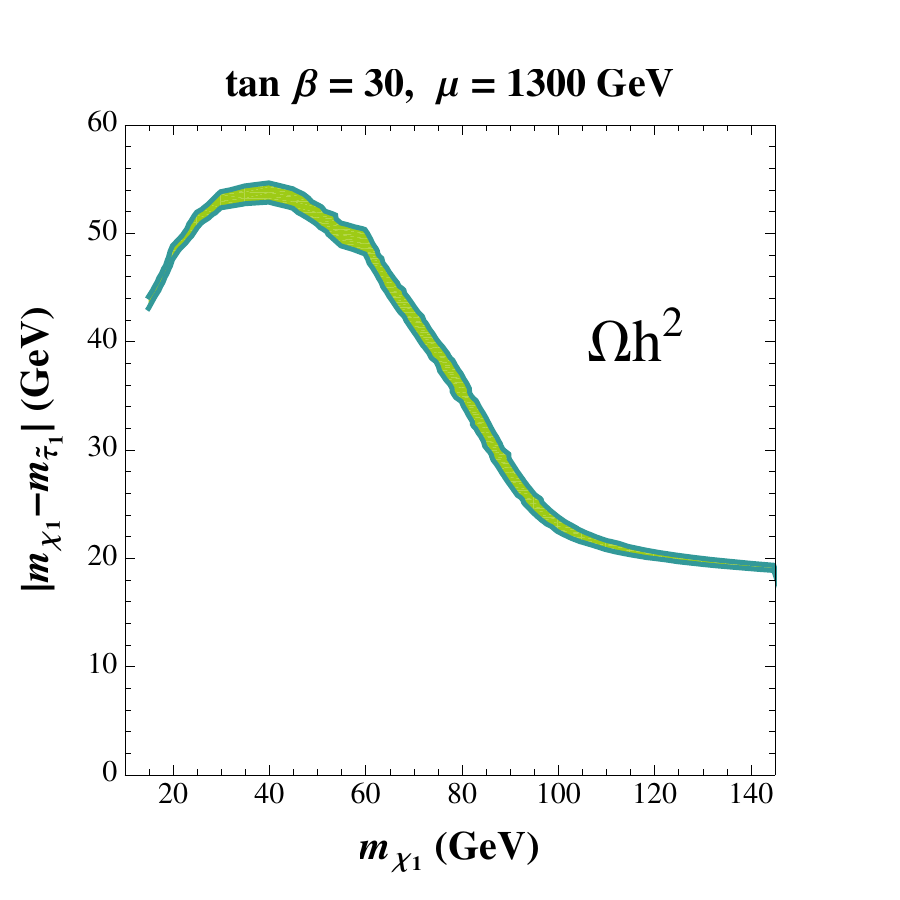}  &
\includegraphics[width=0.45\textwidth]{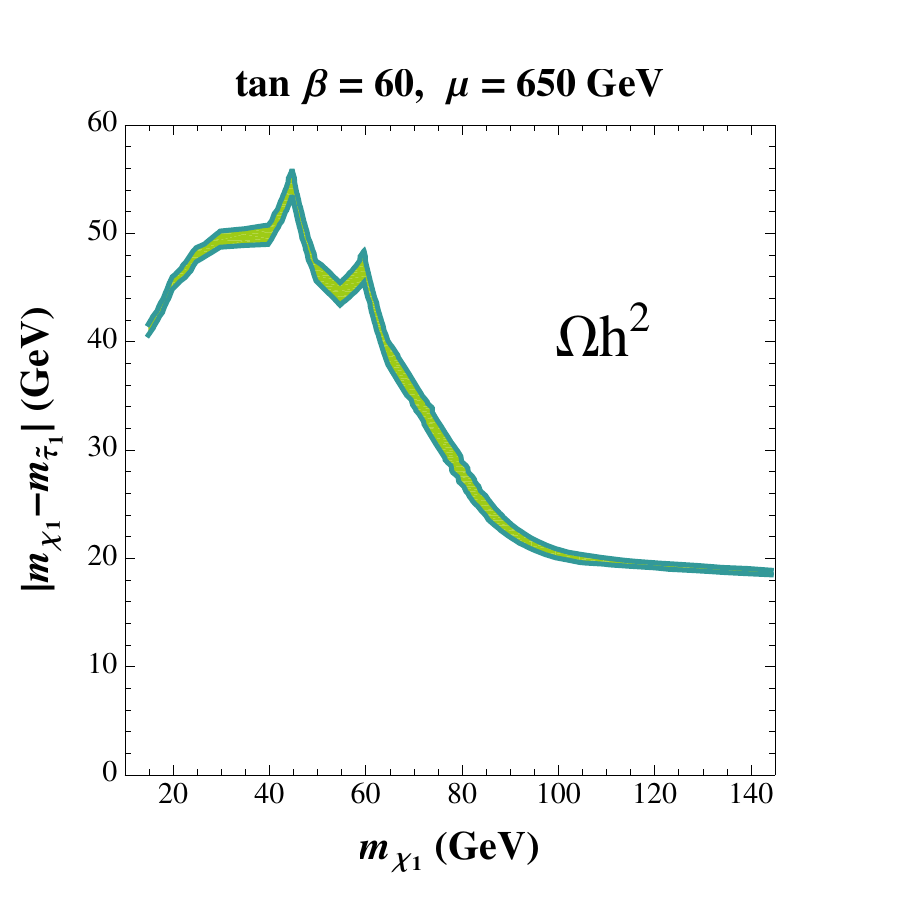}  \\
&\\
(c) & (d) \\
\includegraphics[width=0.45\textwidth]{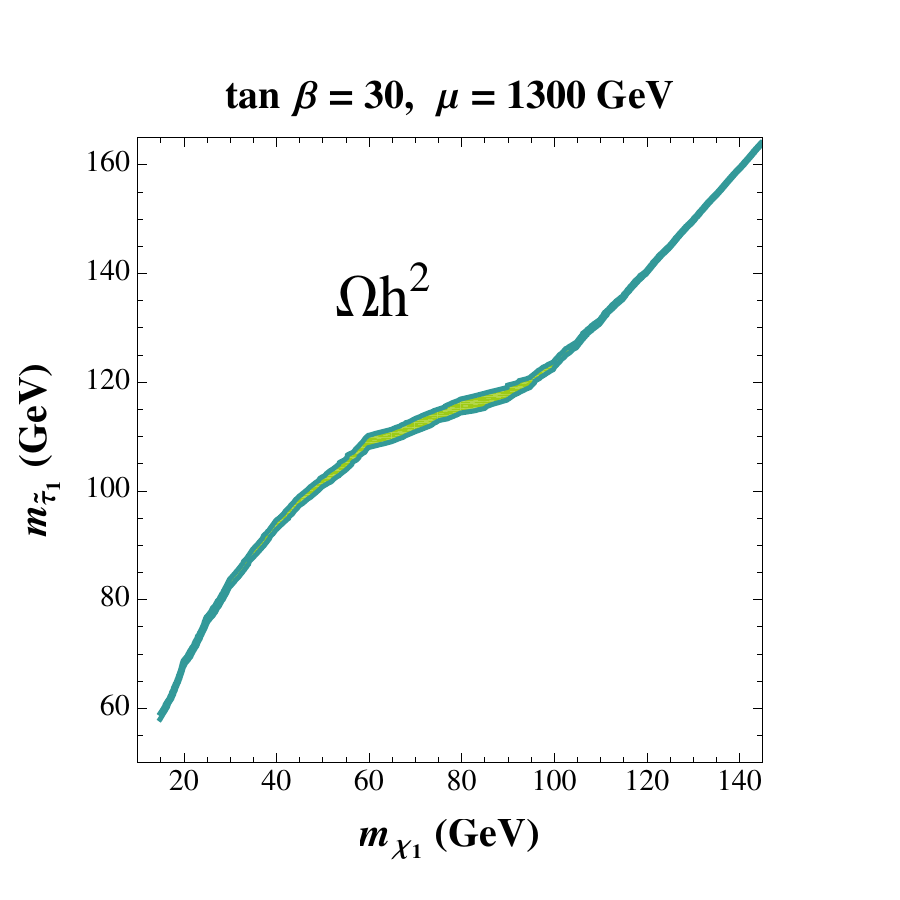}  &
\includegraphics[width=0.45\textwidth]{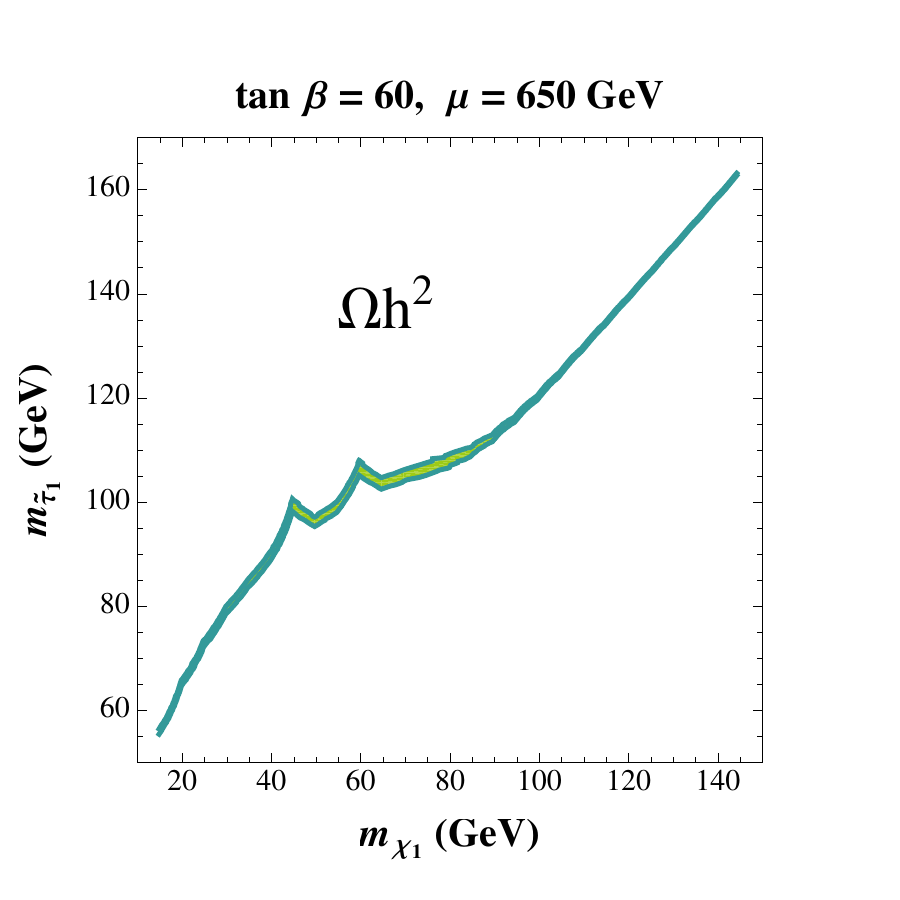}  \\
\end{tabular}
\end{center}
\caption{\textit{(a) \& (b):} Difference between the lightest stau and the lightest neutralino masses, and \textit{(c) \& (d):} lightest stau mass necessary to obtain the observed Dark Matter density as a function of the neutralino mass, for \textit{ (a) \& (c):} $\mu = 1300$~GeV, $\tan\beta = 30$  and \textit{ (b) \& (d):} $\mu = 650$~GeV, $\tan\beta = 60$. The 1st and 2nd generation soft slepton masses are 500 GeV to be consistent with $(g_\mu-2)$. The relevant squark parameters are:  $m_{Q_3}=m_{u_3}=850$~GeV, $A_t=1.4$ TeV. All other parameters are 2 TeV, a part from $M_2=400$ GeV.}
\label{darkmatterstaudel}
\end{figure}

An interesting possibility in the light stau scenario is the generation of the proper dark matter relic density through the light stau co-annihilation with the LSP, generically considered to be the light neutralino ($\chi_1$)~\cite{Edsjo:1997bg},\cite{Ellis:1999mm}.  In order to compute these effects we have used the public programs  \texttt{DarkSUSY}\cite{Belanger:2010pz}
 and \texttt{MicrOMEGAs}\cite{Gondolo:2004sc},  which give consistent results. 

The top two plots in Fig.~\ref{darkmatterstaudel} show the dependence of the mass difference between the neutralino and the stau as a function of the neutralino mass to get the correct relic abundance. In the bottom panels of Fig.~\ref{darkmatterstaudel} the same results are represented in the stau-neutralino mass planes. Light staus are obtained by varying the left- and right-handed stau soft supersymmetry-breaking masses, keeping $m_{L_3} \simeq m_{e_3}$.  We look at two examples keeping $\mu \tan\beta$ a constant, but with different values of $\mu$ and $\tan\beta$. 

The mass difference between the stau and the neutralino parametrizes the strength of the co-annihilation contribution. Light staus can co-annihilate with neutralinos leading to a neutral gauge boson,  $(Z/\gamma)$, and a $\tau$ lepton in the final state.  In the region of parameters under study, the stau is relatively strongly coupled to the Higgs and therefore the coannihilation into a light Higgs and a $\tau$ through the t-channel exchange of a stau becomes very relevant,
\begin{equation}
\chi_1 \; \tilde{\tau}_1 \rightarrow h \; \tau,
\end{equation} 
and in fact turns out to be the dominant channel for most of the range of light neutralino masses under consideration. Since in our analysis we fixed a sizable value of $M_2$ (400 GeV) and we have relatively large values of $\mu$, $\chi_1$ will tend to be mostly a bino. Therefore, the $\tau\tilde{ \tau}_1\chi_1$ coupling is dominated by the hypercharge coupling,  but receives small modifications depending on the values of $\mu$, $M_2$ and $\tan\beta$.  Hence, the amplitude of this annihilation channel is approximately proportional to $\mu\tan\beta$ due to the $\tilde{\tau}_1\tilde{\tau}_1 h$ coupling, but will have small variations with the explicit values of $\mu$ and $\tan \beta$ individually~(we have kept $M_2$ the same for all the plots). This is clearly seen by noting that the maximum mass difference between the stau and the neutralino decreases by about 5 GeV when comparing Fig.~\ref{darkmatterstaudel} (a) and (b), where $\mu \tan \beta$ is constant.  For relatively large values of the neutralino and stau masses, for both  plots, a proper Dark Matter density requires mass differences of the order of 20~GeV. As the neutralino mass goes below about 80~GeV, the co-annihilation cross section starts to increase due to the proximity of the energy threshold for the production of Higgs and tau.  At some point, the neutralino and stau masses become sufficiently small, $m_{\tilde{\chi}_1}\lesssim 30$~GeV and $m_{\tilde{\tau}_1}\lesssim 90$~GeV, so that the Higgs and tau can no longer be produced by annihilation of the stau and the neutralino. At this point the mass difference starts decreasing from a maximum value of order 50--60 GeV at the threshold energy for Higgs and tau production.

There are also the Higgs and $Z$ induced s-channel contributions that grow for smaller values of $\mu$ due to the increase of the Higgsino component of the neutralino, and lead to the presence of small peaks in the mass difference, as is most apparent in Fig.~\ref{darkmatterstaudel} (b). 

It is clear from Figs.~\ref{darkmatterstaudel} (c) and (d) that for stau masses close to the LEP limit, the neutralino mass is of order 30 to 40 GeV, associated with the mass difference of about 50 to 60 GeV mentioned above. Additionally we checked that for large values of the stau soft supersymmetry-breaking parameters, $m_{L_3}$ and $m_{e_3}$, for which larger values of $\mu$ are necessary for the same $\tan\beta$ to obtain a stau mass $\sim$ 90 GeV,  the stau-neutralino mass difference required to get the proper relic density increases by a few GeV, but the results remain qualitatively the same.  Hence, in general, a proper relic density may be obtained in the region where the diphoton Higgs decay width is enhanced, for stau masses close to the LEP limit and neutralino masses of about 30 to 40~GeV.

\section{RG Evolution to High Energies}
\label{RGE}

As we have discussed in the previous sections, the values of the third generation squarks and sleptons are constrained by the requirement of a 125~GeV Higgs with an enhanced diphoton decay rate. 
Although most of the interesting low energy physics properties are governed by the coupling of the light staus to the Higgs, which is proportional to $ \mu\tan\beta$, the renormalization group (RG) evolution of the soft supersymmetry-breaking parameters~\cite{Martin:1993zk}  depends strongly on the $\tau$ Yukawa coupling and weakly on the exact value of $\mu$.  Large values of $\tan\beta$ lead to  large $\tau$ and bottom Yukawa couplings. The effect of strong Yukawas has been extensively discussed in the literature (see, for instance~\cite{Carena:1993bs}--\cite{Blazek:2002ta}).  They tend to suppress the value of the scalar masses at low energies, and therefore large RG effects on the slepton soft supersymmetry-breaking masses are expected in their evolution from the messenger scale.  These effects would be stronger for large values of the messenger scale, close to the GUT scale, and become weaker for values of the messenger scale of order $10^5$~GeV.  It is important to stress that the hypercharge D-term contributions to the RG evolution are also important, in particular due to the large values of the Higgs and squark soft supersymmetry-breaking masses compared to the slepton ones.  Moreover, the $\tau$-Yukawa coupling effects depend on the value of the $H_d$ soft supersymmetry-breaking square mass parameter, which at large values of $\tan\beta$ tends to increase with energy due to the large bottom Yukawa effects.   


One may also require that the first and second generation sleptons are in the range consistent with a relevant contribution to $(g_\mu-2)$.  In the previous section, we have seen that an explanation of the observed anomalous magnetic moment of the muon may be obtained by assuming left-handed smuon masses that are of order 500~GeV, only somewhat larger than the third generation ones. 

If, in order to suppress dangerous flavor effects, we demand flavor independence of the soft supersymmetry-breaking parameters at the messenger scale, we can obtain relevant information on the value of the messenger scale. In order to do that, we shall take a bottom-up approach, noting  that the first and second generations have negligible Yukawas and therefore small RG running, governed by the weak gaugino masses and couplings.  At large $\tan\beta \simeq 60$ and  relatively large values of the stop masses and mixing, the CP-odd Higgs mass and the Higgsino mass parameters required in this scenario, flavor universality can only be obtained in a natural way for low values of the messenger scale. Such low values of the messenger scale are  associated with light gravitinos, for which our previous computation of the Dark Matter density would be invalid.


 \begin{figure}
\begin{center}
\begin{tabular}{cc}
\textbf{(a)}: $M \simeq 10^7$~GeV, $\tan\beta=60$& \textbf{(b):} $M \simeq  10^{16}$~GeV\\
\\
\includegraphics[width=0.48\textwidth]{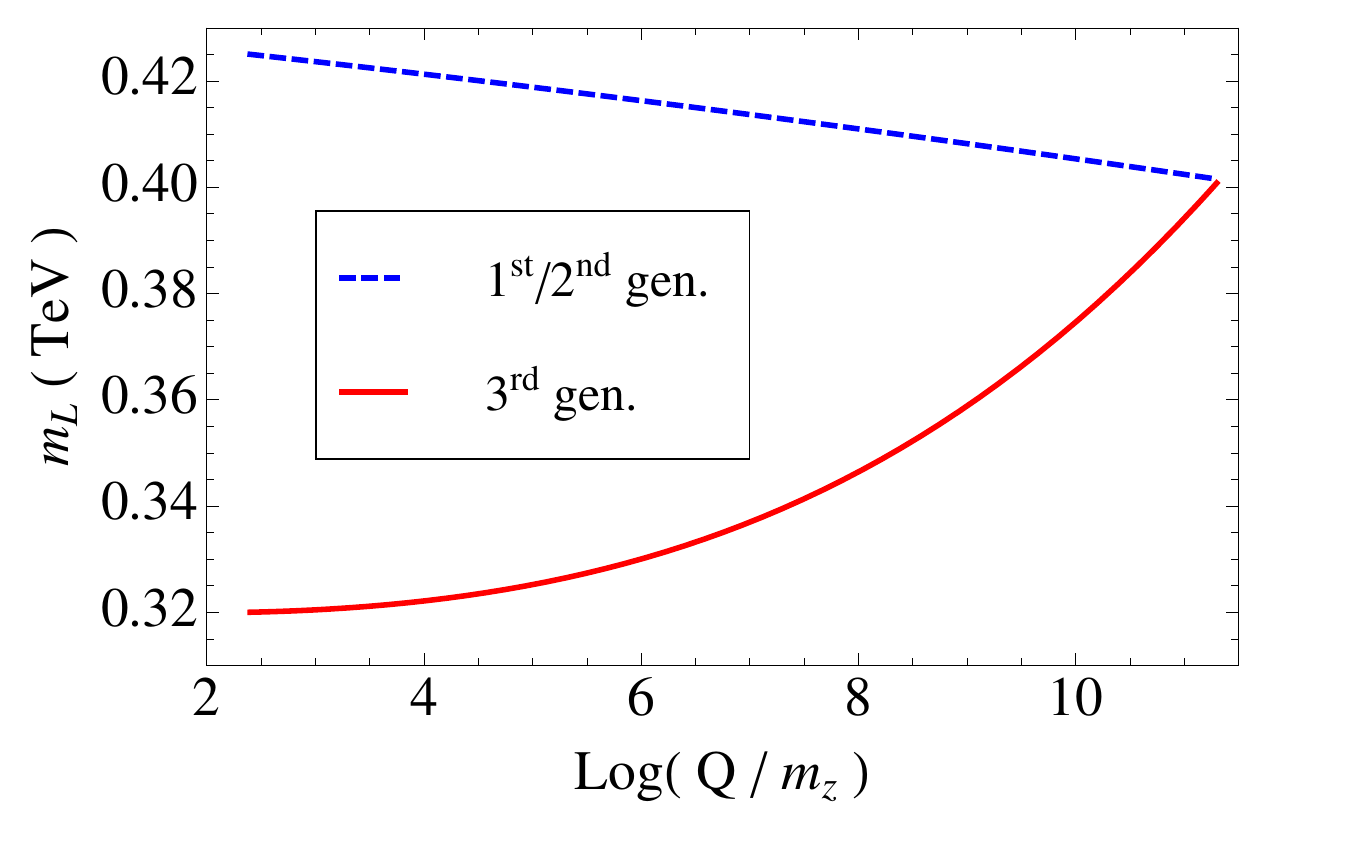} &
\includegraphics[width=0.48\textwidth]{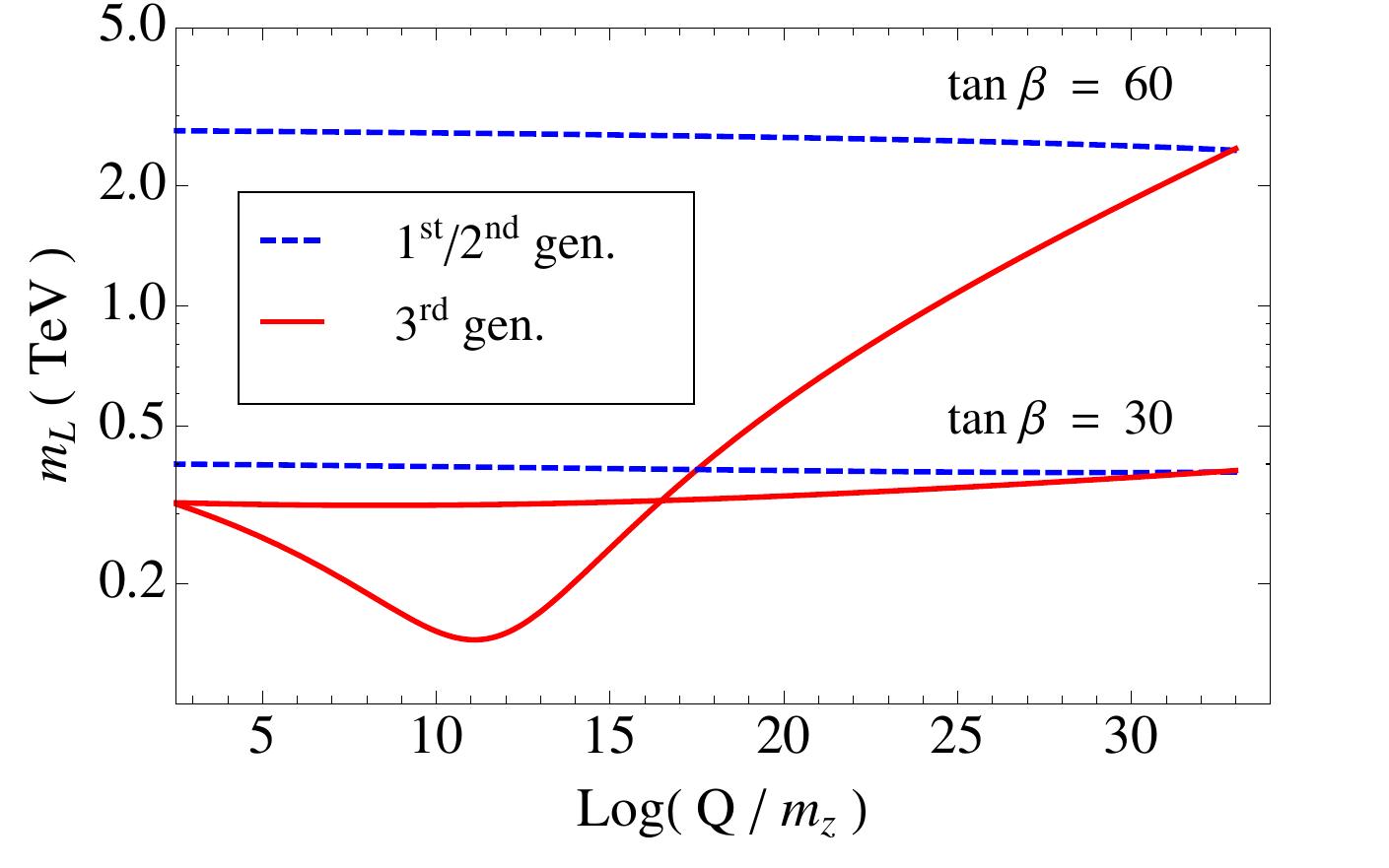}  \\
\end{tabular}
\end{center}
\caption{\footnotesize{$m_{L}$ evolution from the TeV to the Messenger scale, $M$. Blue: 1st/2nd generation, Red: 3rd generation.  $\tan \beta =60$ is associated with $\mu=650$ GeV and $\tan \beta=30$ with $\mu=1300$ GeV.}}
\label{runningmL}
\end{figure}

The stau effects on the Higgs spectrum and properties depend only on the product  $\mu \tan\beta$. One can soften the RG evolution of the slepton mass parameters while keeping the Higgs properties intact by decreasing the value of $\tan\beta$  and simultaneously increasing the value of $\mu$.  Fig.~\ref{runningmL}  shows the evolution of the left-handed slepton soft supersymmetry-breaking parameters as one raises the energy towards the GUT scale, for low energy values of the left-handed stau soft breaking parameter $m_{L_3} \simeq 320$~GeV~(the evolution of the other soft supersymmetry-breaking parameters is shown in Appendix A).  Values of $\tan \beta =60$ are associated with $\mu=650$ GeV and $\tan \beta=30$ with $\mu=1300$ GeV. The previously described properties are clear in these figures. For $\tan \beta=60$, flavor independent values close to the ones necessary to obtain the muon anomalous magnetic moment can only be naturally obtained for relatively small messenger scales $M$ (of order $10^7$~GeV for the example given in Fig.\ref{runningmL}), as can be seen by comparing the RG evolution in Fig.~\ref{runningmL} (a) and (b) for $\tan\beta=60$. However, for $\tan\beta=30$, the $\tau$ Yukawa effects become significantly weaker and the weak scale slepton masses consistent with $(g_\mu-2)$ can acquire flavor independent values at the GUT scale, shown in Fig.~\ref{runningmL} (b). Moreover, as we have shown in the previous section, a proper Dark Matter relic density can be also obtained for these conditions.  

Let us stress in closing that we have not performed a detailed scanning of the parameter space consistent with a 125~GeV Higgs  with enhanced diphoton decay widths. Neither have we considered variations of the finite threshold corrections to the bottom Yukawa coupling at low energies~\cite{deltamb}--\cite{deltamb2}. Modifications of the specific values of the squark and Higgs mass parameters, and of the gluino mass, may induce relevant changes in the running of the different parameters of the theory. Therefore,  one cannot exclude the possibility of obtaining flavor independent parameters at energies of the order of the GUT scale even for large values of $\tan\beta\sim 60$. Finally, the most relevant flavor violation effects are related to the first and second generation squark and slepton mass parameters. Therefore even if the first and second generation masses do not unify with the third generation at the messenger scale, there is still the possibility of avoiding problems in the flavor sector.


\section{Probing the Light Stau Scenario at the LHC }

Light staus, with masses of order a 100~GeV and large mixing are interesting since they predict the presence of a light sneutrino and an additional heavier stau. Therefore, although the staus are only produced weakly, the collider signatures associated with this scenario may have a very complex and rich structure.

In this work we will focus on the  direct weak production of staus (and tau sneutrinos) through an s-channel exchange of $Z$ (or $W$) gauge bosons. We give an estimate of the discovery reach at the LHC for both 8 TeV and 14 TeV center of mass energies. These channels turn out to be quite model independent, since they depend only on the masses and mixings of staus (and sneutrinos) and would be open even in the scenario of very heavy squarks and gluinos. As shown in the second column of Table~\ref{tab:channels}, the typical signature will be multi-taus, missing energy and weak gauge bosons, giving rise to additional leptons.

In our analysis, we used parton level results obtained from \texttt{Madgraph 5}~\cite{Alwall:2011uj}. We emphasize that a more realistic simulation  would necessarily include parton showering, hadronization, and detector simulation.  A properly matched matrix element plus parton shower simulation can be particularly important for the estimation of $W+$jets background. However, our simplified analysis is suitable for  our goal of  obtaining a rough order of magnitude estimate of the discovery reach. 

\subsection{Status of Current LHC Stau Searches}

At present, the ATLAS collaboration is investigating the presence of third generation sleptons produced through cascade decays. They analyze final states containing taus, leptons, hard jets and large missing energy, arising from (relatively light) squarks/gluinos decaying directly or through cascades into the $\tilde\tau$ NLSP~\cite{Aad:2012rt,ATLAS:2012ag}. This channel is complementary to the ones we investigate, but is more model dependent. 

On the other hand, final states similar to the ones we are interested in have been already investigated by CMS~\cite{Chatrchyan:2012ye} in the context of searches for charginos and neutralinos. However, comparing the cross sections listed in the third column of Table~\ref{tab:channels} to the CMS results, we note that the CMS multilepton searches are still not sensitive to our scenario~\footnote{The most promising channel ($\tilde\tau_1\tilde\nu_\tau$ production) would produce at most only $\sim$ 4 events at the 5 fb$^{-1}$ 7 TeV LHC. This rate is below the CMS uncertainty on the number of expected events in the two taus/one lepton channel (see their Table 2)~\cite{Chatrchyan:2012ye}.}. For this reason the most stringent constraint on the mass of the staus is still given by the LEP bound that is around (85-90) GeV for the case of the split stau-neutralino spectrum~\cite{LEPlimit}. In the following, we will propose search strategies which are optimized to enhance the sensitivity to the particular light stau scenario considered in this paper.

\subsection{Weakly Produced Staus}

We propose searches for the direct production of staus, with 
\begin{equation}
\quad\;\;\;\;\;\tilde{\tau}_1 \to \chi_1 \tau\; , \;\;\;\mbox{or}\;\;\;\tilde{\tau}_1 \to \tilde{G} \tau\;.
\end{equation}
 
Dark Matter relic density, associated with large messenger scales and hence a neutralino DM, tends to predict a large mass difference between the stau and the DM candidate (see Sec.~\ref{sec:DM}). Alternatively, we could have a low messenger scale and a very light gravitino. In both cases,  the missing energy tends to be sizable, which could facilitate searches for light staus. To simplify our presentation, we choose $m_{\rm LSP} = 35 $ GeV, as preferred by the neutralino LSP scenario.  We have checked that lowering the neutralino mass does not significantly alter our conclusions.

Possible channels to look for stau and sneutrino direct production are shown in Table~\ref{tab:channels}. In particular, we show the possible signatures of several channels at the LHC and the production cross sections for an example point in parameter space where $m_{L_3}=m_{e_3}=280$ GeV, $\tan\beta=60$, $\mu=650$ GeV and $M_1=35$ GeV, giving a light stau, $m_{\tilde\tau_1}\sim95$ GeV, a very light LSP, $m_{\chi_1}\sim 35$ GeV and a light sneutrino, $m_{\tilde\nu_\tau}\sim 270$ GeV. Typically, at the 8 TeV LHC, we expect cross sections of the order of tens of fb only for the $\tilde\tau_1\tilde\tau_1$ and $\tilde\tau_1\tilde\nu_\tau$ channels.

\begin{table}[h,t]
\begin{center}
\begin{tabular}{|c|c|c|c|}
\hline
\hline
                                             & Signature                   & 8 TeV LHC (fb) & 14 TeV LHC (fb)\\
\hline
\hline
$pp\rightarrow\tilde\tau_1\tilde \tau_1$     & $2\tau, E\!\!\!\!/_T$       & 55.3 & 124.6\\
\hline
$pp\rightarrow\tilde\tau_1\tilde \tau_2$     & $2\tau,Z, E\!\!\!\!/_T$     & 1.0 & 3.2\\
\hline
$pp\rightarrow\tilde\tau_2\tilde \tau_2$     & $ 2\tau,2Z, E\!\!\!\!/_T$   & 0.15 & 0.6\\
\hline
$pp\rightarrow\tilde\tau_1\tilde \nu_\tau$   & $2\tau, W, E\!\!\!\!/_T$    & 14.3 & 38.8\\
\hline
$pp\rightarrow\tilde\tau_2\tilde \nu_\tau$   & $2\tau, W,Z, E\!\!\!\!/_T$  & 0.9 & 3.1\\
\hline
$pp\rightarrow\tilde\nu_\tau\tilde \nu_\tau$ & $2\tau, 2W, E\!\!\!\!/_T$   & 1.6& 5.3\\
\hline
\hline
\end{tabular}
\end{center}
\caption{Possible stau and sneutrino direct production channels with their signatures at the LHC. The cross sections shown are computed for $m_{L_3}=m_{e_3}=280$ GeV, $\tan\beta=60$, $\mu=650$ GeV and $M_1=35$ GeV.}
\label{tab:channels}
\end{table}

The most promising channel seems to be $pp\rightarrow\tilde\tau_1\tilde \nu_\tau$ because of the additional $W$ boson in the final state.  More specifically, for the relatively large mass difference between the sneutrino and the lightest stau obtained in the region consistent with an enhanced diphoton rate,  the dominant production and decay mode is expected to be
\begin{equation}
p p \to \tilde{\tau}_1 \tilde{\nu}_\tau  \to \tilde\tau_1 (W\tilde\tau_1)\to \tau \chi_1 W \tau \chi_1\;.
\end{equation}
The final state is two hadronic taus, missing energy and the $W$ decaying leptonically, which leads to a much cleaner signal than the $\tilde\tau_1\tilde\tau_1$ production. The competing mode would be the direct decay of the sneutrino into a neutrino and a neutralino, which, however, tends to have a smaller branching ratio due to the relative smallness of the hypercharge gauge coupling. In the following, we shall concentrate on this channel at the 8 TeV LHC.

The main physical background contributing to the $2\tau + W + E\!\!\!\!/_T$ signature is given by $W + Z/ \gamma^*$, with a cross section of 900 fb at the 8 TeV LHC. We also need to include the $W+$ jets background with jets faking taus in our study.

We generate events for the signal, physical background and fake background requiring taus (jets) with a $p_T$ threshold, $p_T^{\tau (j)}>10$ GeV, $\Delta R>0.4$ and $|\eta|<2.5$. 
We demand two loose $\tau$-tags: the efficiency of the boosted decision tree (BDT) hadronic tau identification is about $60\%$, independent of $p_T$, while
achieving a jet background rejection factor of 20 - 50~\cite{ATLAS:tau}. 
The cross sections for signal and backgrounds associated with these requirements are given in the second column of Table~\ref{tab:basic}\footnote{For the $W+$ jets background we generated events with up to 4 jets in the final state. In the table we are presenting the sum of $Wjj$, $Wjjj$ and $Wjjjj$ backgrounds.}.

\begin{table}[t]
\begin{center}
\begin{tabular}{|c|c|c|c|}
\hline
\hline
                                                              & Total (fb) & Basic (fb)  & Hard Tau (fb) \\
\hline
\hline
Signal     & 0.6 & 0.16 &0.07 \\
\hline
Physical background, $W+Z/\gamma^*$     & 15 & 0.25& $\lesssim 10^{-3}$ \\
\hline
$W+$ jets background          &$4\times 10^3$ & 26 & 0.3 \\
\hline
\hline
\end{tabular}
\end{center}
\caption{Cross sections for the signal and the physical and fake backgrounds after $\tau$-tags at the 8 TeV LHC: after imposing acceptance cuts  $p_T^{\tau (j)}>10$ GeV, $\Delta R>0.4$ and and $|\eta|<2.5$ (second column); with the additional requirement $p_T^{\ell}>70$ GeV and $E\!\!\!\!/_T>70$ (third column); imposing that the $\tau$ is not too boosted $p_T^{\tau}<75$ GeV (fourth column).}
\label{tab:basic}
\end{table}

\begin{table}[t]
\begin{center}
\begin{tabular}{|c|c|c|c|}
\hline
\hline
                                                              & Total (fb) & Basic (fb)  & Hard Tau (fb) \\
\hline
\hline
Signal     & 1.6 & 0.26 &0.11 \\
\hline
Physical background, $W+Z/\gamma^*$     & 27 & 0.32& $\lesssim 10^{-3}$ \\
\hline
$W+$ jets background          &$10^4$ & 39 & 0.25 \\
\hline
\hline
\end{tabular}
\end{center}
\caption{Cross sections for the signal and the physical and fake background after $\tau$-tags at the 14 TeV LHC: after imposing $p_T^{\tau (j)}>10$ GeV, $\Delta R>0.4$ and and $|\eta|<2.5$ (second column); with the additional requirement $p_T^{\ell}>85$ GeV and $E\!\!\!\!/_T>85$ (third column); imposing that the $\tau$ is not too boosted $p_T^{\tau}<80$ GeV (fourth column).}
\label{tab:basic14}
\end{table}

Due to the sizable mass splitting between the sneutrino and the stau, the lepton coming from the $W$ decay in the signal is expected to be more boosted than the one from background. For this reason, strong cuts on the $p_T$ of the lepton and on the missing energy can significantly improve the signal over background ratio. In the third column of Table~\ref{tab:basic}, labeled ``Basic'', we show our results after imposing $p_T^{\ell}>70$ GeV and $E\!\!\!\!/_T>70$ GeV.  As we can see from the table, this set of basic cuts can efficiently suppress the $W+Z/\gamma^*$ background to a rate comparable to the one of the signal. In addition, we note that the two taus coming from the physical background are typically expected to have an invariant mass close to the $Z$ peak.  Therefore, a veto of the $\tau_1\tau_2$ invariant mass close to $m_Z$ will further suppress the physical background. However, given our stringent cuts of $p_T^{\ell}$ and $E\!\!\!\!/_T$ (and the further cut on the $p_T$ of the leading $\tau$ presented below), we notice that the additional improvement from $Z$-veto is marginal.  Since our signal is statistics limited, we choose not to further impose this cut in our study. On the other hand, in a fully realistic study, one could certainly include $Z$-veto as a possible variable to be optimized together with other cuts. 

\begin{figure}[t!]
\begin{center}
\includegraphics[width=0.58\textwidth]{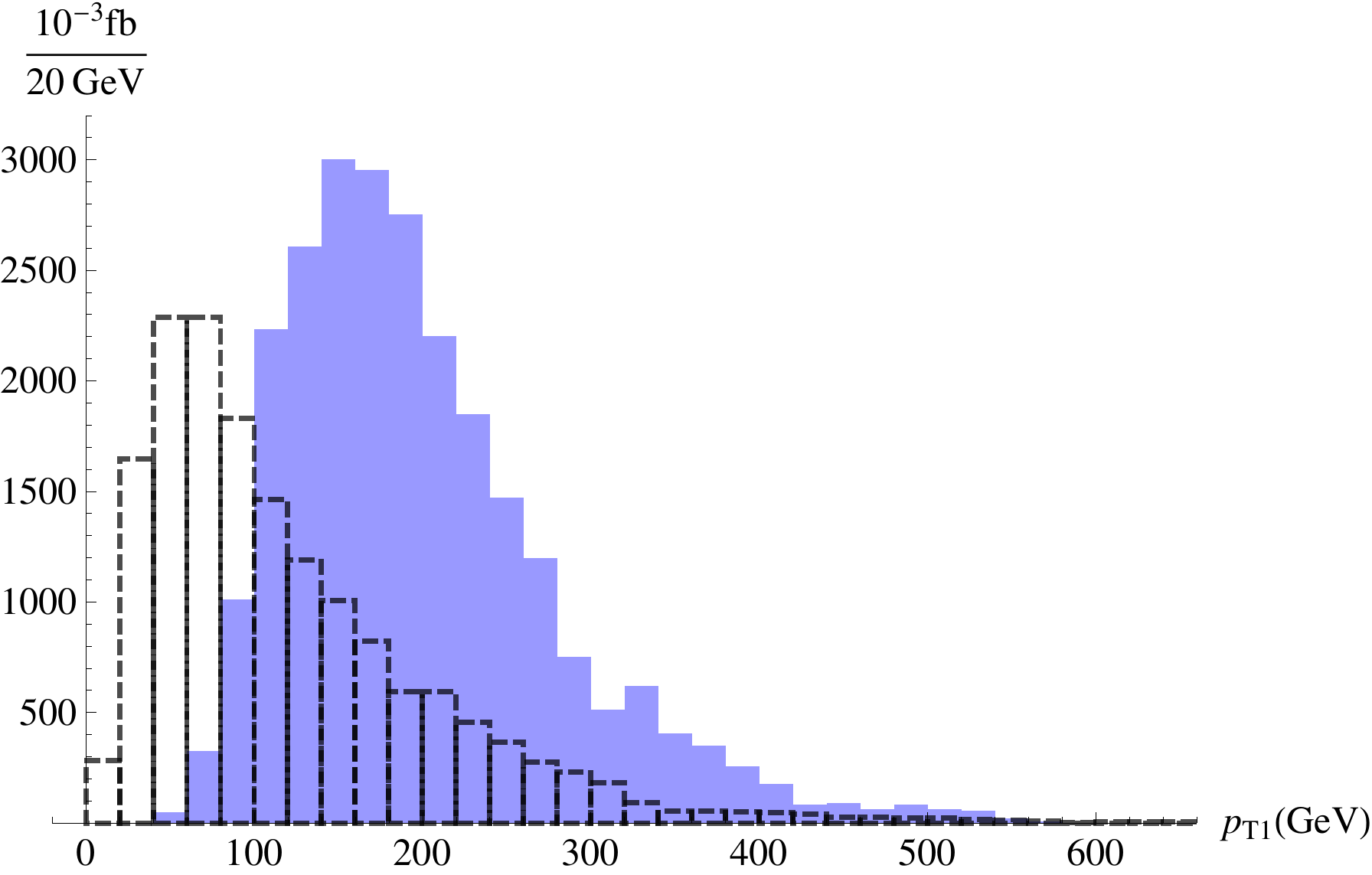} 
\end{center}
\caption{\footnotesize{$p_T$ distribution for the leading jet faking a tau of the $W+$ jets background (in blue) and for the leading tau of the signal (black dashed) at the 8 TeV LHC. The events shown satisfy the basic set of cuts ($p_T^{\ell}>70$ GeV and $E\!\!\!\!/_T>70$ GeV). The signal has been scaled by a factor of 100 for visibility.}}
\label{distributions}
\end{figure}
The $W+$jets background is still significant at this stage. As shown by the blue distribution in Fig.~\ref{distributions}, the leading fake tau will recoil against the lepton and hence will also be rather hard. On the other hand, in the signal process, the $\tilde{\tau}_1$ only receives a small boost even if it is one of the decay products of the $\tilde{\nu}$. The $p_T$ of the leading tau is always largely determined by $m_{\tilde{\tau}_1} - m_{\rm LSP}$ and remains sufficiently soft (see black dashed distribution in Fig.~\ref{distributions}). Consequently, a veto on hard $\tau$s can reduce the fake background, while keeping the signal almost unchanged. In the fourth column of Table~\ref{tab:basic}, labeled ``Hard Tau'', we show our results for signal and backgrounds, after requiring the leading $\tau$ to have $p_T^{\tau_1}<75$ GeV. Due to this veto on hard taus, signal and (fake) background are approximately the same order of magnitude. 

In spite of low statistics, 
we believe that this channel deserves attention, especially in view of the possible 200 fb$^{-1}$ of luminosity  expected from the 14 TeV LHC run\footnote{ Note that reducing the mass of the sneutrino sizably increases the direct production cross section of sneutrino - stau pairs. However, the mass splitting between the sneutrino and the stau would decrease, reducing the boost of the $W$ boson coming from the sneutrino decay. Therefore lighter sneutrinos will not necessarily enhance the LHC reach for the $\tilde\nu_\tau\tilde\tau_1$ channel.}

In Table~\ref{tab:basic14} we present the cross sections for the signal, physical background and the $W+$ jets background for the 14 TeV LHC with a set of cuts very similar to the ones used for the 8 TeV LHC: the requirement on the $p_T$ of the lepton and on the missing energy are slightly more demanding, $p_T^\ell>85$ GeV and $E\!\!\!\!/_T>85$, and the veto on hard taus has been slightly relaxed, $p_T^{\tau}<80$ GeV. From the numbers in Table~\ref{tab:basic14} we see that the ratio between signal and (fake) background is of $\mathcal O(1)$ and that one can expect tens of signal events with 200 fb$^{-1}$ of luminosity.

We would like to briefly discuss the $\tilde\tau_1\tilde\tau_1$ channel. As shown in Table~\ref{tab:channels}, the total production cross section for two staus is a factor of four larger than the direct production cross section of a stau and a sneutrino at the 8 TeV LHC. However, the present double hadronic $\tau$ trigger is rather demanding: the $p_T$ thresholds are 29 GeV and 20 GeV for the leading and sub-leading hadronic $\tau$s~\cite{ATLAS:ditau}. Imposing this basic requirement to  trigger and asking for two loose taus decreases the cross section of the $\tilde\tau_1\tilde\tau_1$ channel from the 55.3 fb presented in Table~\ref{tab:channels} to 7 fb at the 8 TeV LHC. 

The main sources of physical backgrounds are $Z+Z/\gamma^*$, and $W^+ W^-$. A veto on the invariant mass of the $\tau_1\tau_2$ system close to the $Z$ peak helps  in considerably reducing the $Z+Z/\gamma^*$ physical background. In particular, we checked that demanding the invariant mass to be outside the interval $70\,{\rm{GeV}}<m_{\tau_1\tau_2}<130$ GeV~\cite{Chatrchyan:2012vp}, reduces the $Z+Z/\gamma^*$ physical background to 0.4 fb while keeping the signal still at 4 fb.  The $W^+ W^-$ background is however still significant after the $Z$-veto: 27 fb. We could further reduce the $W^+ W^-$ background by noticing that most of the taus from the $W$ decay have $p_T^\tau < m_W /2$. Imposing $p_T^{\tau_{\rm 1, 2}} > 50$ GeV and $E\!\!\!\!/_T>80 $ GeV brings the $W^+ W^-$ background down to about 0.3 fb, about the same as the signal (0.4 fb) after these cuts.

However, the real challenge for this channel is the background from jets faking taus. Such fake background is  dominated by $W+1$ jet, which has the jet faking a tau and the $W$ decaying to an additional tau. In comparison, $Z+$ jets, where the jets fake taus and the $Z$ decays invisibly, is subdominant.  Even after the set of cuts mentioned above,
the signal over background ratio is still $\sim$1$\%$ (due to the large $W+1$ jet background: 57fb), with signal and fake background having very similar distributions for the kinematical observables. For this reason, we believe that the $\tilde\tau_1\tilde\nu_\tau$ is a more promising channel than $\tilde\tau_1\tilde\tau_1$. However, the latter channel could have room for further improvements. An enhancement of jet rejection power could significantly improve the sensitivity. In addition, polarization of the two final $\tau$s could be a very important discriminant between the $\tilde\tau_1\tilde\tau_1$ signal and the $W+1$ jet background~\cite{Cao:2003tr}. Further enhancement of the signal cross sections quoted above may be achieved by considering the contributions coming from $b$-quark annihilation and gluon fusion to the production cross section of staus, as recently shown in Ref.~\cite{Lindert:2011td}.

Finally, an additional very interesting channel is $pp\to\tilde\tau_2\tilde\tau_1\to h \tilde\tau_1\tilde\tau_1$ since it would directly probe the coupling between the Higgs and the two staus entering in the Higgs to diphoton rate. However, as shown in Table~\ref{tab:channels}, the cross section is rather small for this channel to be relevant at the LHC.

\section{Conclusions}

Recent Higgs searches are consistent with the production of a light Higgs, with a mass  of about 125~GeV, and a somewhat enhanced  $\sigma(gg \to h) BR(h \to \gamma\gamma)$. No such enhancement is observed in other Higgs production channels, suggesting that this effect can at least in part be due to an increase of the Higgs decay branching ratio into photons.  

Within the MSSM, light staus, close to the LEP limit of $\sim$~85-90 GeV, with large mixing may produce an enhancement of the Higgs diphoton rate without affecting the Higgs $ZZ$ rates. In this article, we have studied the phenomenological properties of this light stau scenario. 

We have shown that in general one should expect an increase of $\Delta\rho$, which leads to an enhancement of $m_W$ by 10 to 40 MeV with respect to the value predicted in the SM. Moreover,  values of $(g_{\mu}-2)$ may be obtained for smuon masses slightly larger than, but of the order of, the required stau supersymmetry-breaking parameters.

The RG evolution to high energies demonstrates that to obtain the weak scale parameters consistent with the light stau scenario and flavor universality at  $M_{GUT}$, large values of the soft supersymmetry-breaking parameters, of the order of a few TeV, may be required. However, this requirement maybe softened by lowering $\tan \beta$ or by lowering the messenger scale.

The model is consistent with the observed Dark Matter relic density, provided the neutralino is a few tens of GeV lighter than the light stau. For instance $ 35\,{\rm{GeV}}\lesssim m_{\tilde{\chi}_1} \lesssim 55$~GeV when $90\,{\rm{GeV}}\lesssim m_{\tilde{\tau}_1} \lesssim 100$~GeV. 

Finally, the light stau coupled with a relatively light sneutrino presents distinctive collider signatures. We propose possible strategies to probe this scenario through direct weak production of stau and tau sneutrino. We demonstrate that the associated production of $\tilde{\tau}_1 \tilde{\nu}_\tau$ is within reach at the next run of the LHC at 8 TeV. While this could be the first signal of this scenario, $\tilde{\tau}_1\tilde{\tau}_1$ production  may also prove to be useful with improvement of $\tau$-identification and further optimization of the cuts.  A dedicated study of the search potential is necessary and highly motivated. \\
~\\
~\\
{\bf \Large Acknowledgements}\\
~\\
We thank T. Han and T. LeCompte for useful discussions. Fermilab is operated by Fermi Research Alliance, LLC under Contract No. DE-AC02-07CH11359 with the U.S. Department of Energy. Work at ANL is supported in part by the U.S. Department of Energy~(DOE), Div.~of HEP, Contract DE-AC02-06CH11357. This work was supported in part by the DOE under Task TeV of contract DE-FGO2-96-ER40956. L.T.W. is supported by the NSF under grant PHY-0756966 and the DOE Early Career Award under grant
DE-SC0003930.
\newpage
\appendix
\section*{APPENDIX}
\section{Soft Parameter Evolution}

\begin{figure}
\begin{center}
\begin{tabular}{c c}
\includegraphics[width=0.45\textwidth]{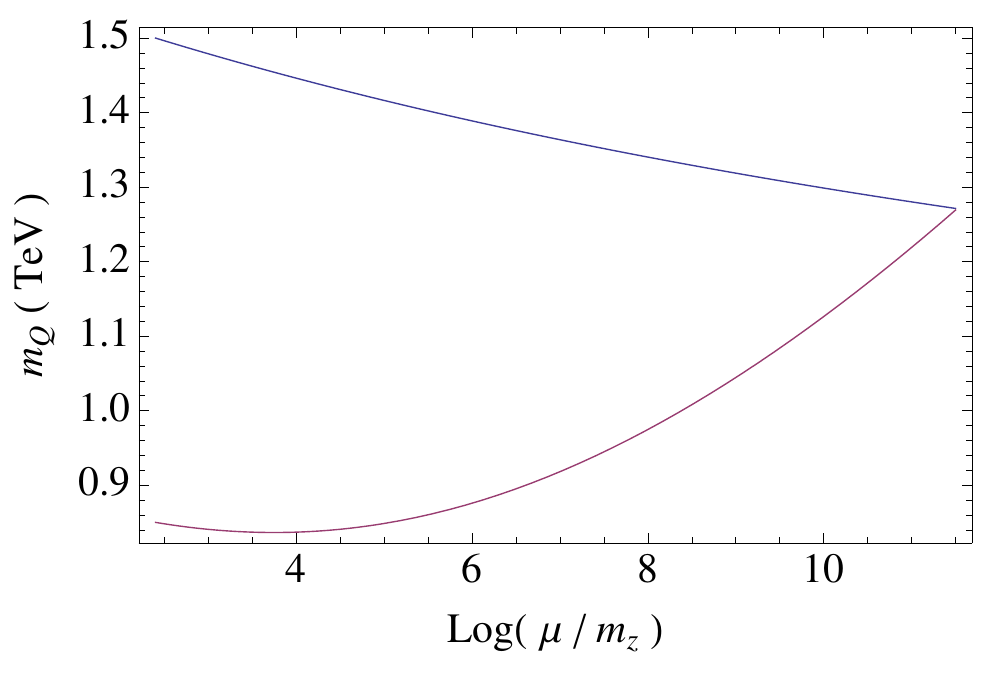}  &
\includegraphics[width=0.45\textwidth]{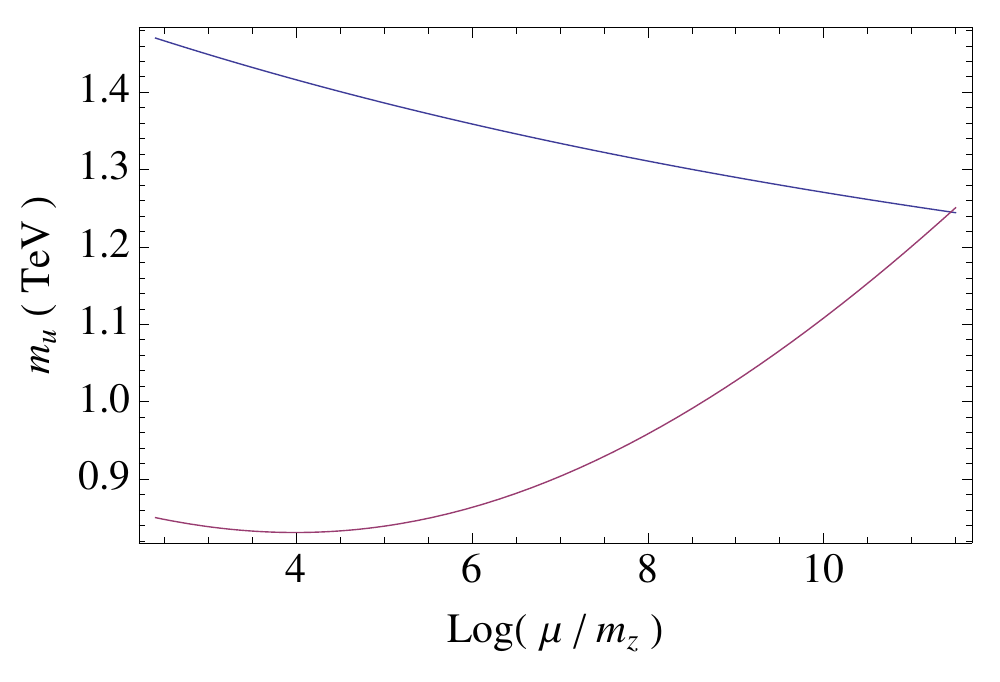}  \\
\includegraphics[width=0.45\textwidth]{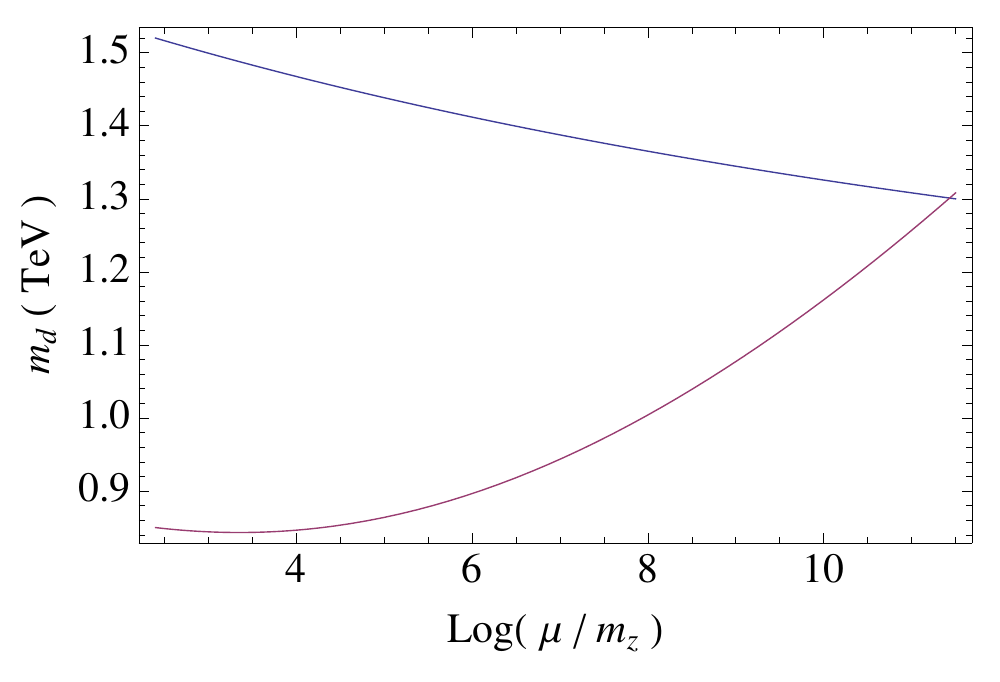}  &
\includegraphics[width=0.45\textwidth]{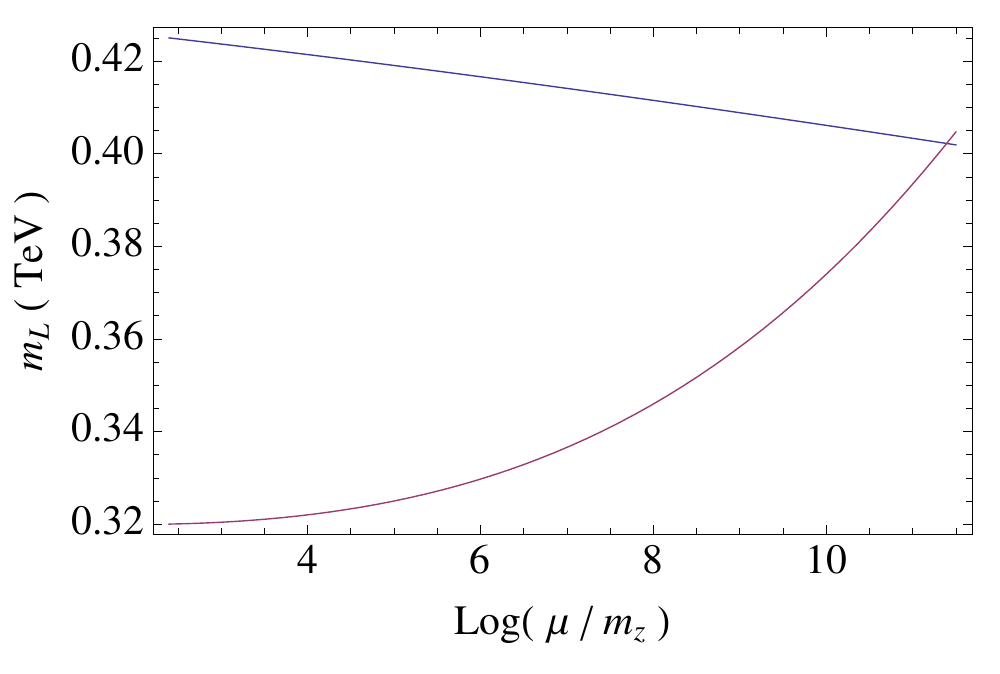}  \\
\includegraphics[width=0.45\textwidth]{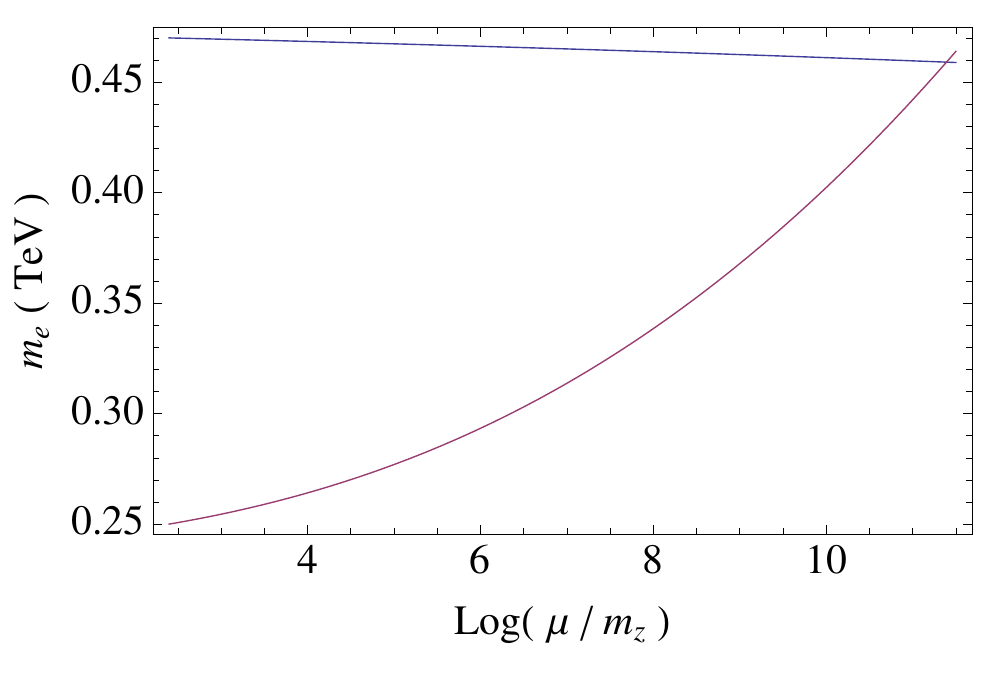}  &
\includegraphics[width=0.45\textwidth]{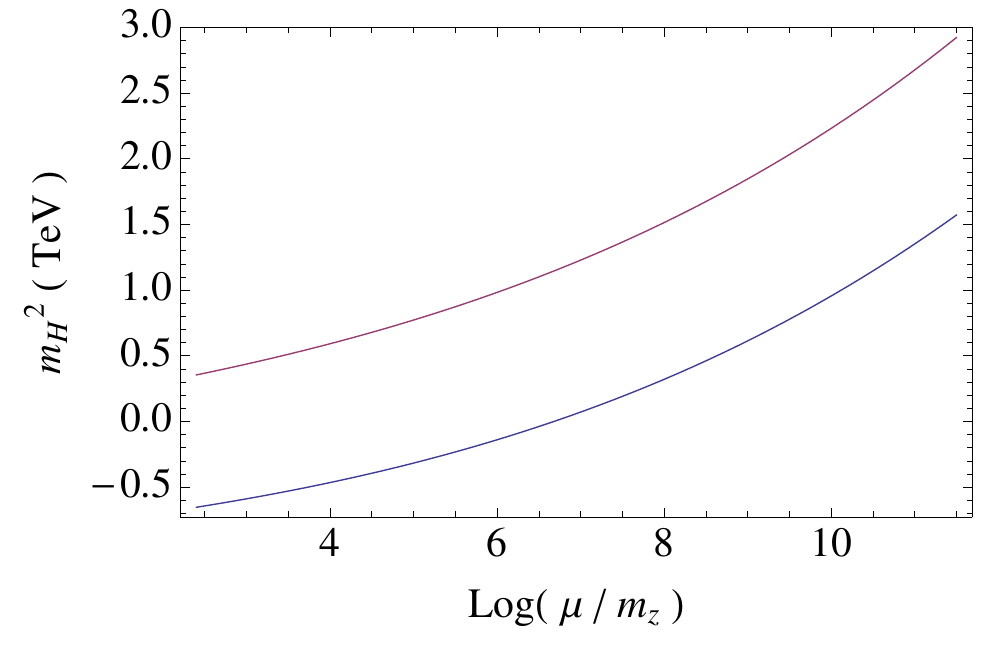} \\
\includegraphics[width=0.45\textwidth]{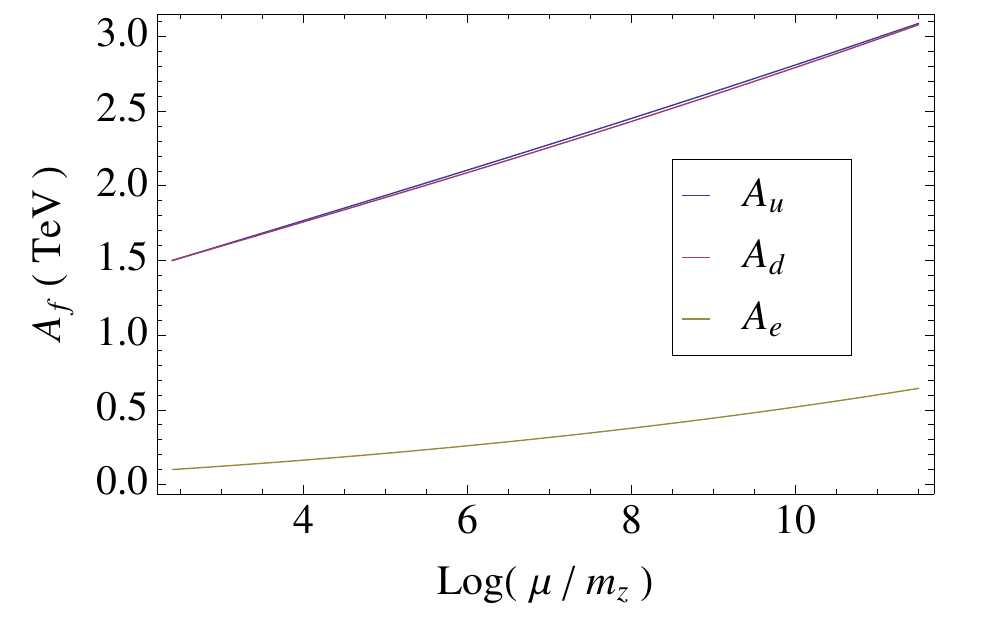}  &
\includegraphics[width=0.45\textwidth]{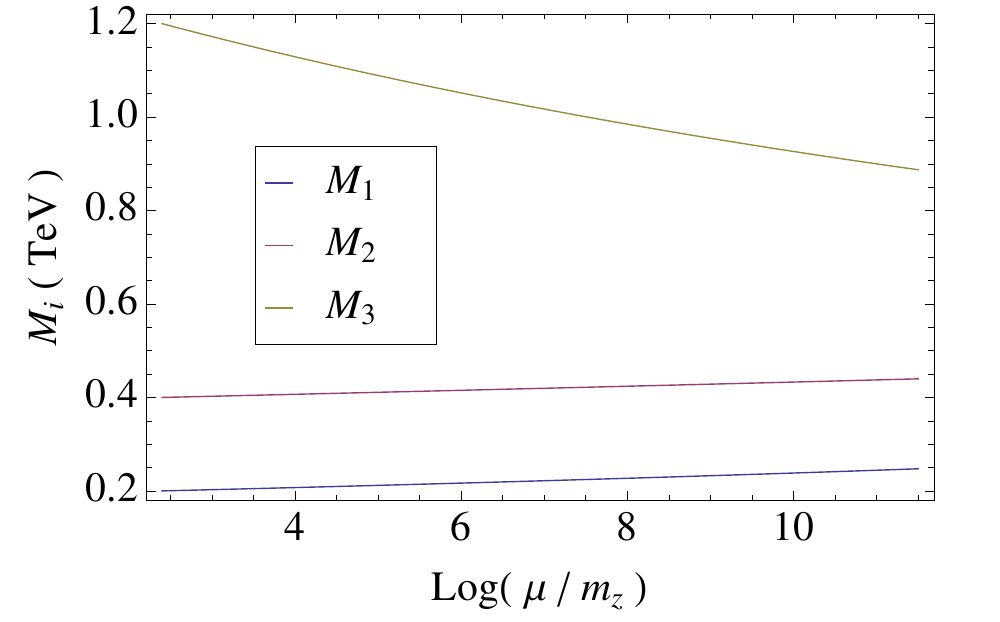}  \\
\end{tabular}
\end{center}
\caption{Mass parameter evolution from the TeV to the Messenger scale $M \simeq  10^7$~GeV for $\tan\beta=60$. }
\label{runningteq95}
\end{figure}

\begin{figure}
\begin{center}
\begin{tabular}{c c}
\includegraphics[width=0.45\textwidth]{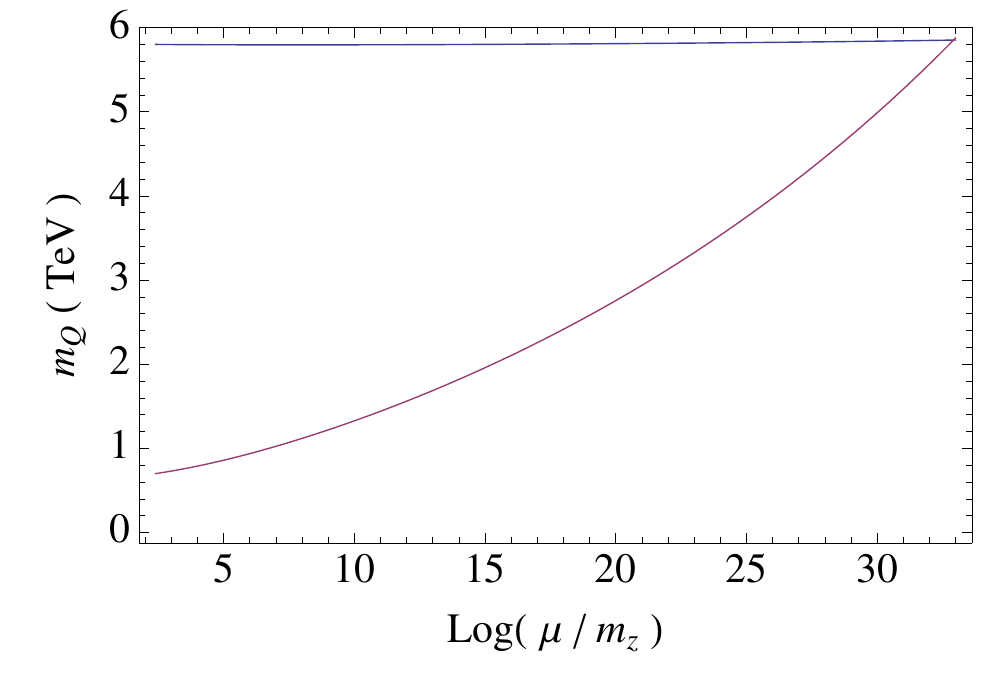}  &
\includegraphics[width=0.45\textwidth]{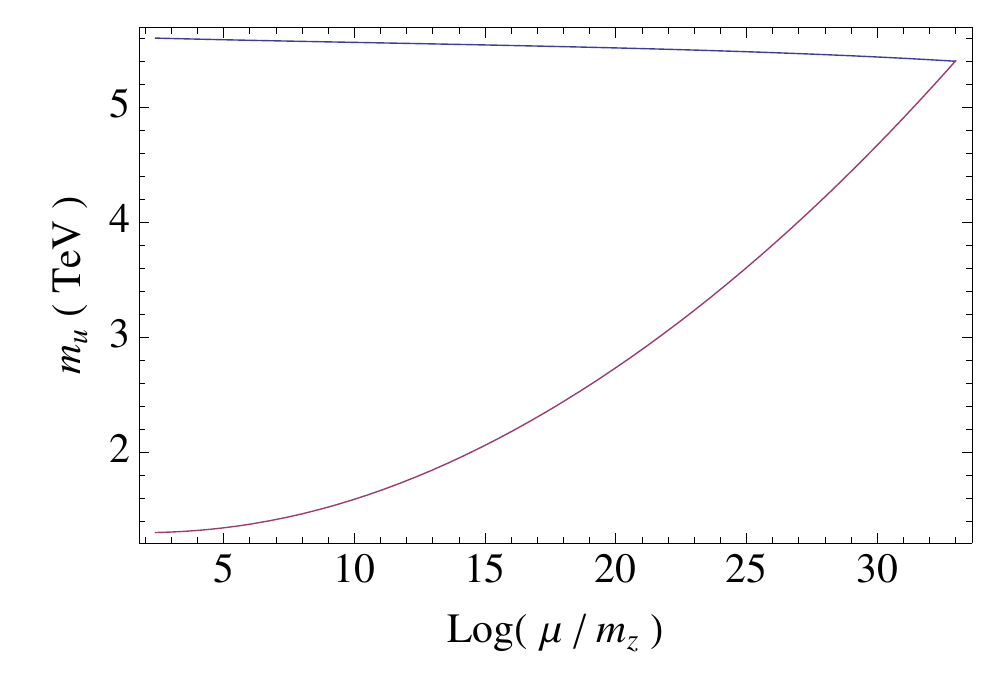}  \\
\includegraphics[width=0.45\textwidth]{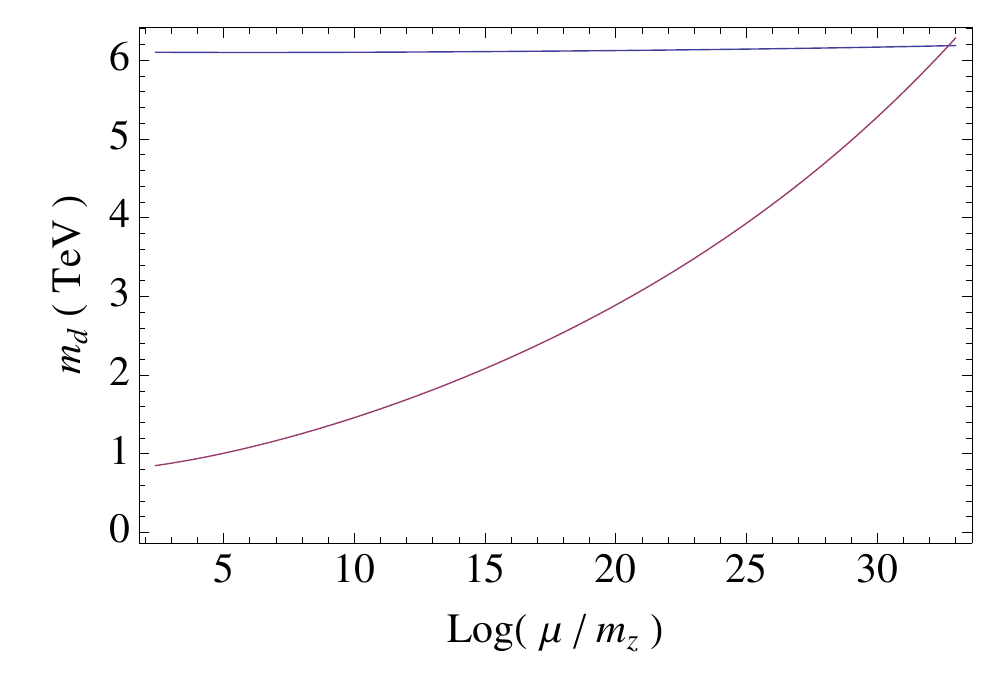}  &
\includegraphics[width=0.45\textwidth]{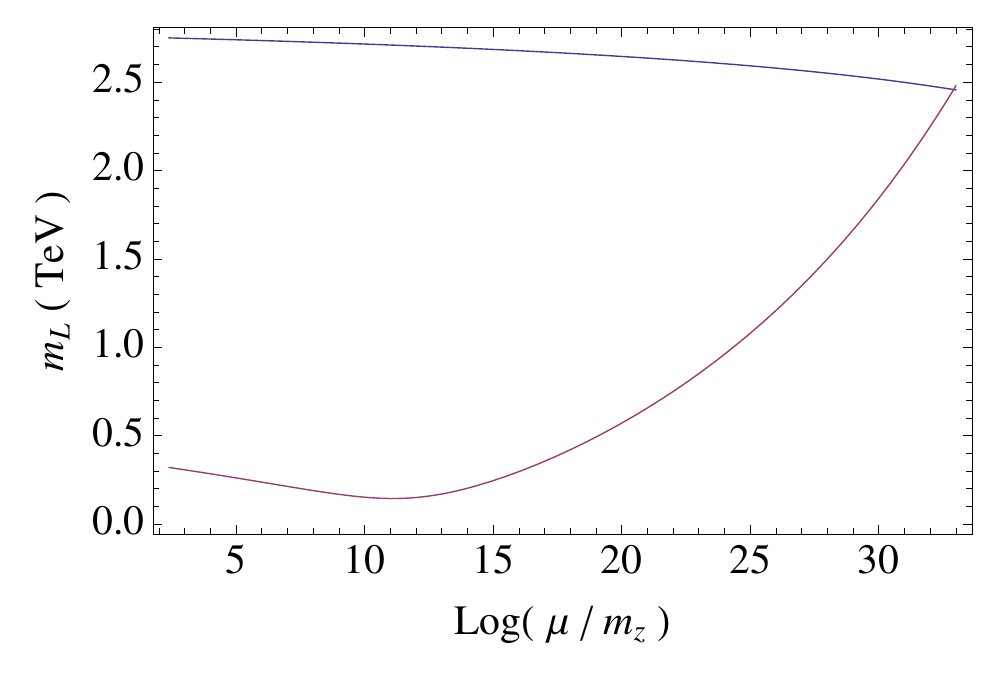}  \\
\includegraphics[width=0.45\textwidth]{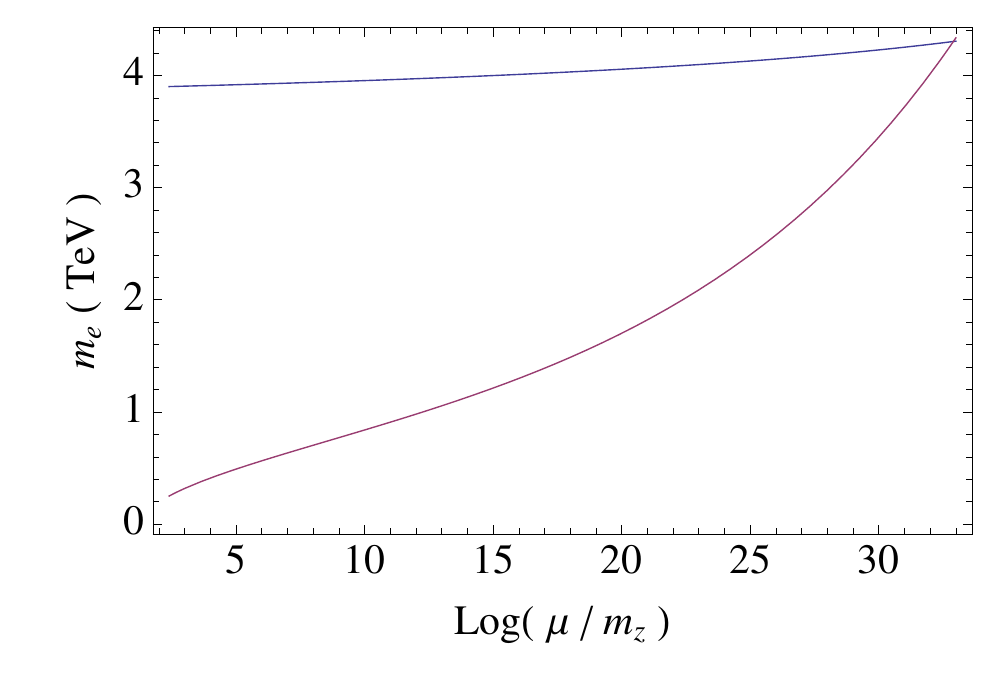}  &
\includegraphics[width=0.45\textwidth]{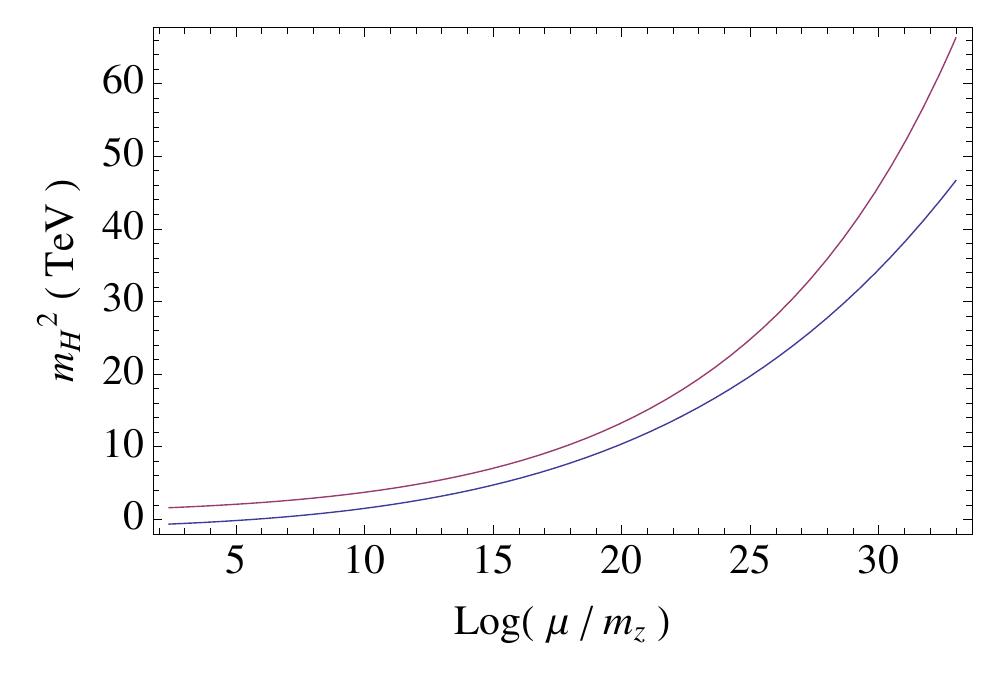} \\
\includegraphics[width=0.45\textwidth]{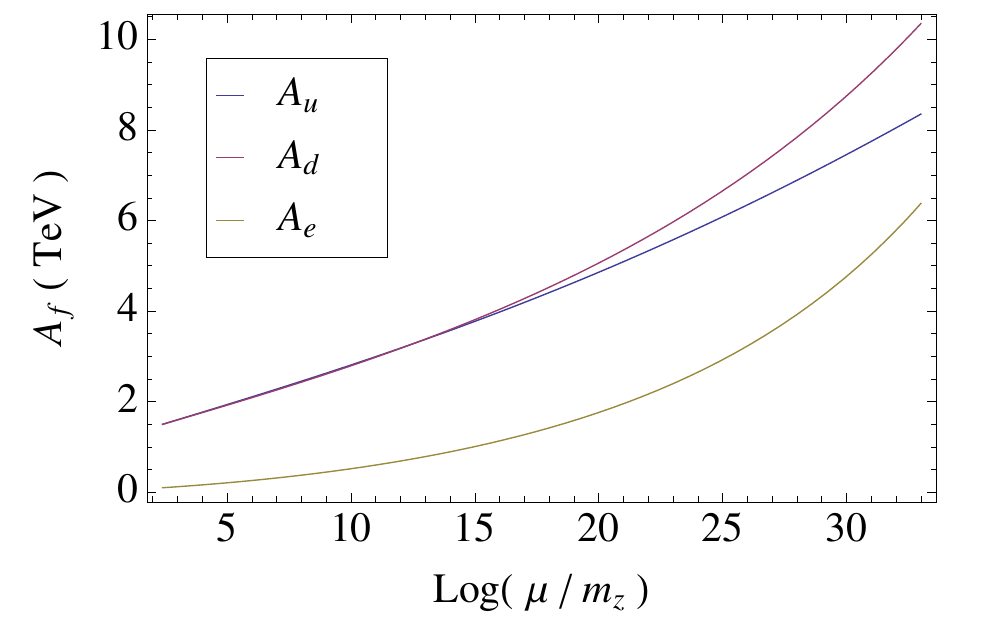}  &
\includegraphics[width=0.45\textwidth]{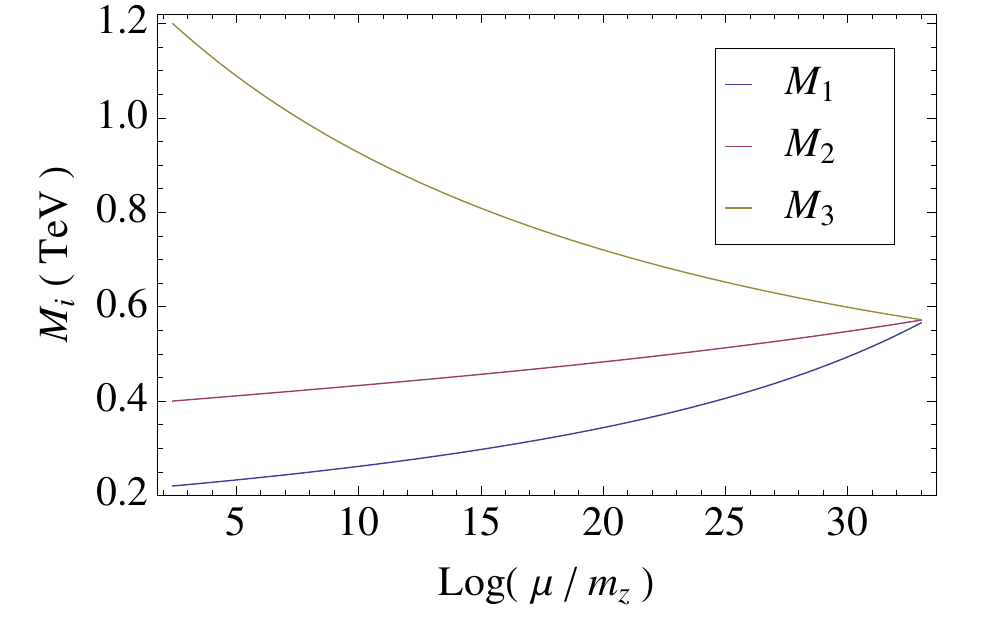}  \\
\end{tabular}
\end{center}
\caption{Mass parameter evolution from the TeV to the Messenger scale $M \simeq  10^{16}$~GeV for $\tan\beta=60$. }
\label{runningteq3360}
\end{figure}

\begin{figure}
\begin{center}
\begin{tabular}{c c}
\includegraphics[width=0.45\textwidth]{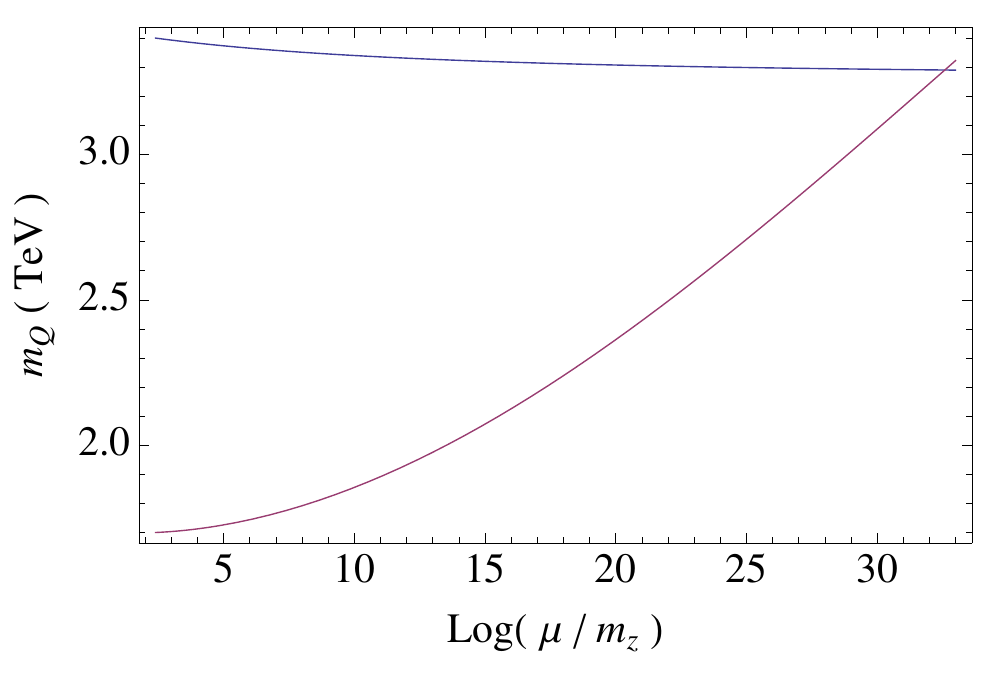}  &
\includegraphics[width=0.45\textwidth]{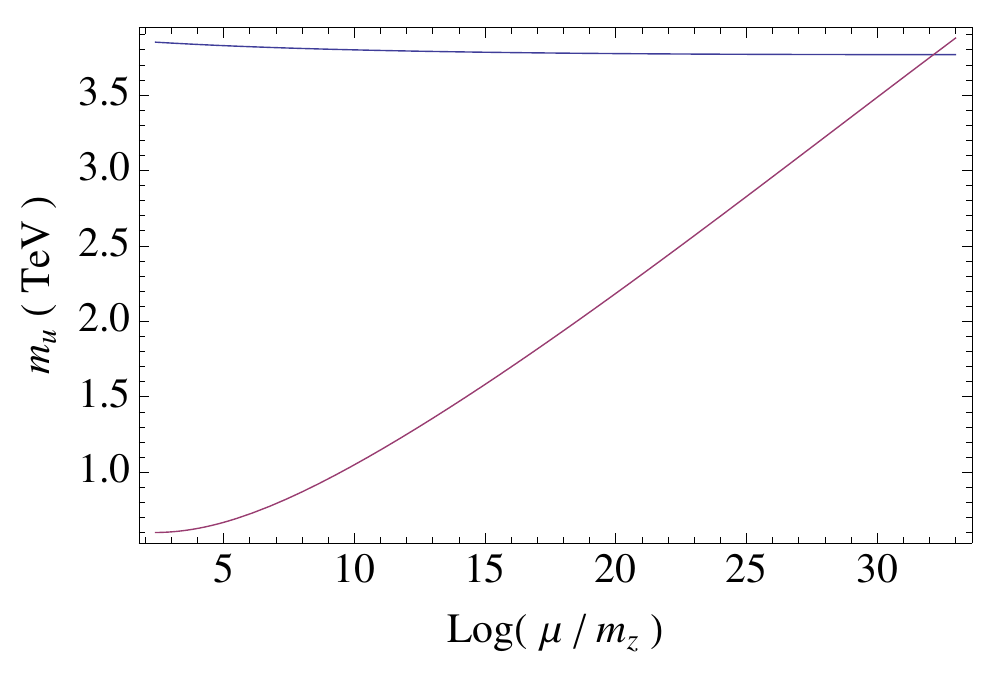}  \\
\includegraphics[width=0.45\textwidth]{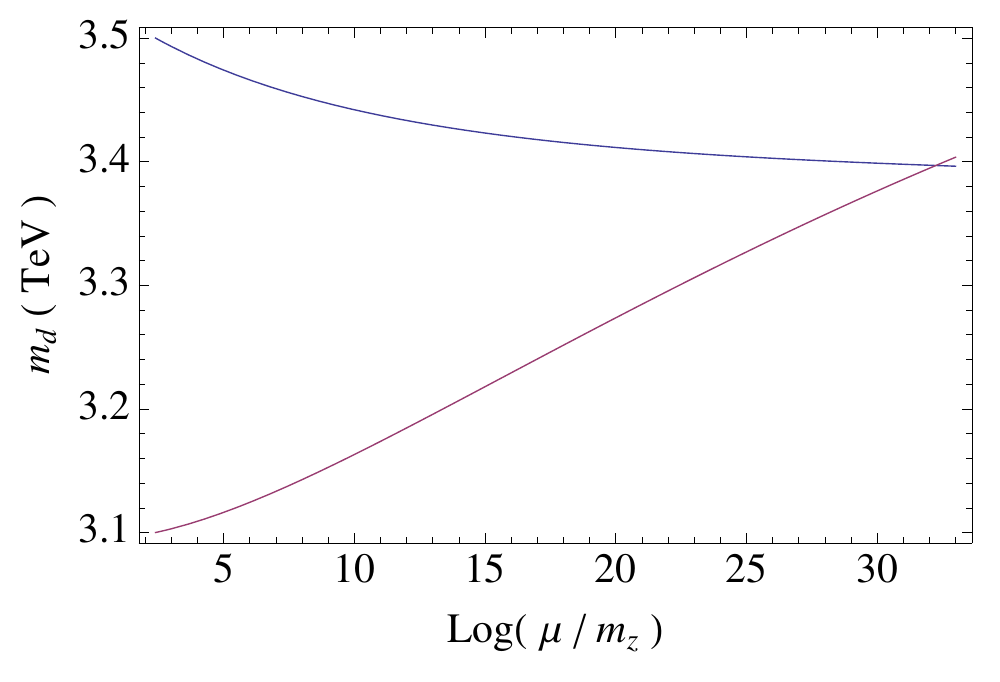}  &
\includegraphics[width=0.45\textwidth]{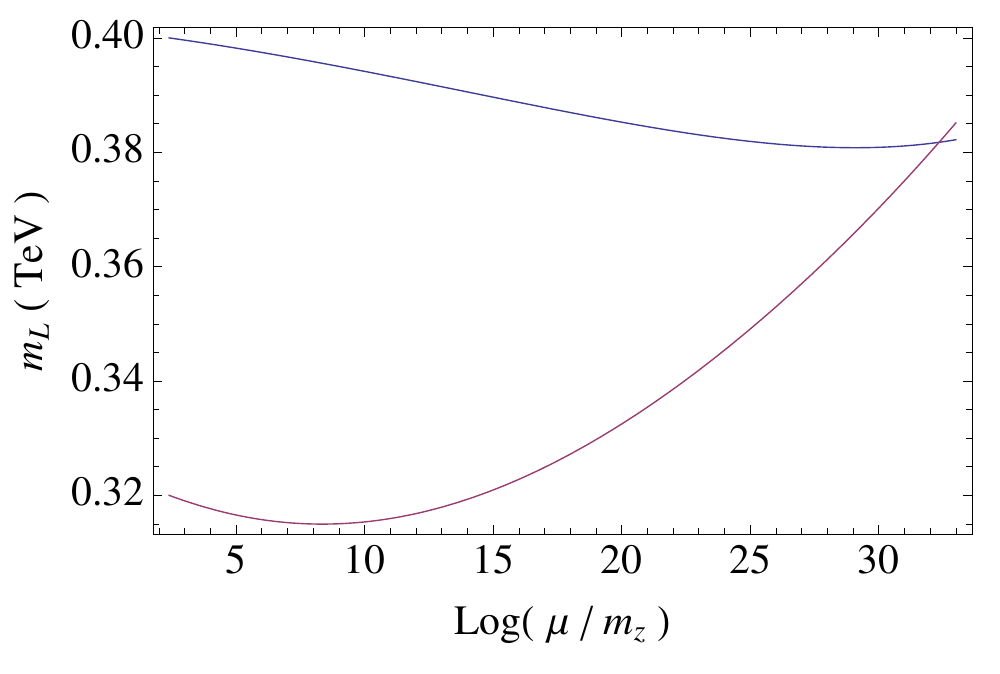}  \\
\includegraphics[width=0.45\textwidth]{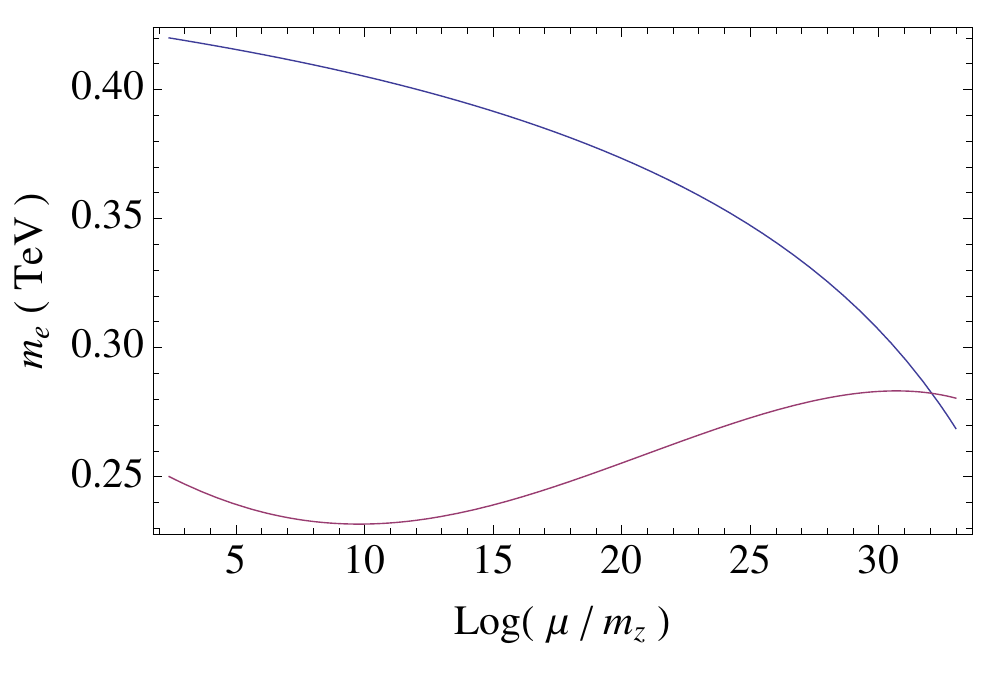}  &
\includegraphics[width=0.45\textwidth]{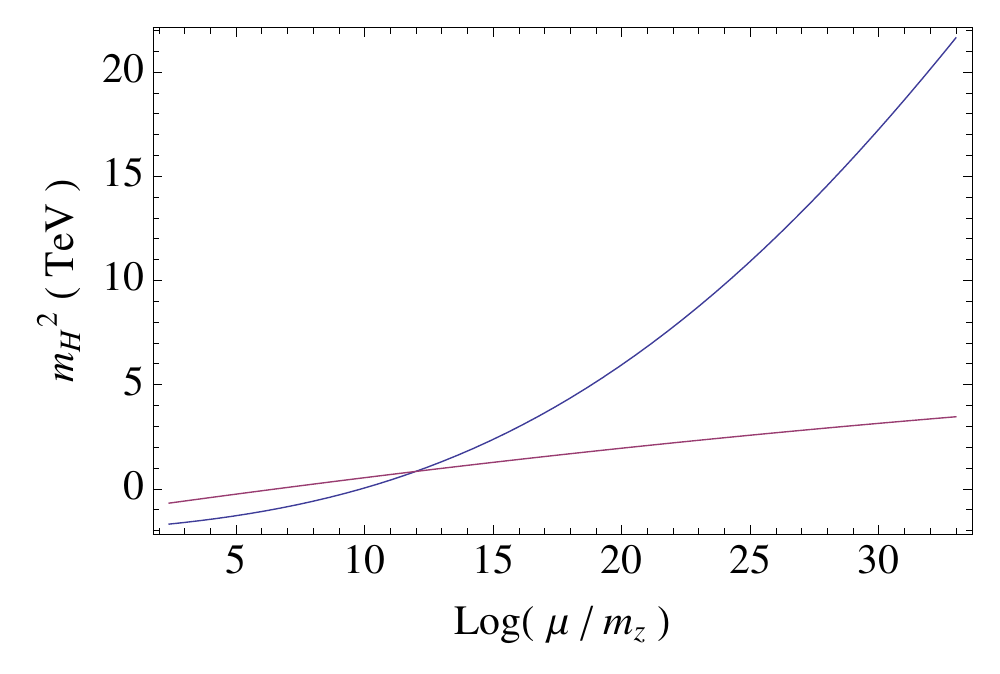} \\
\includegraphics[width=0.45\textwidth]{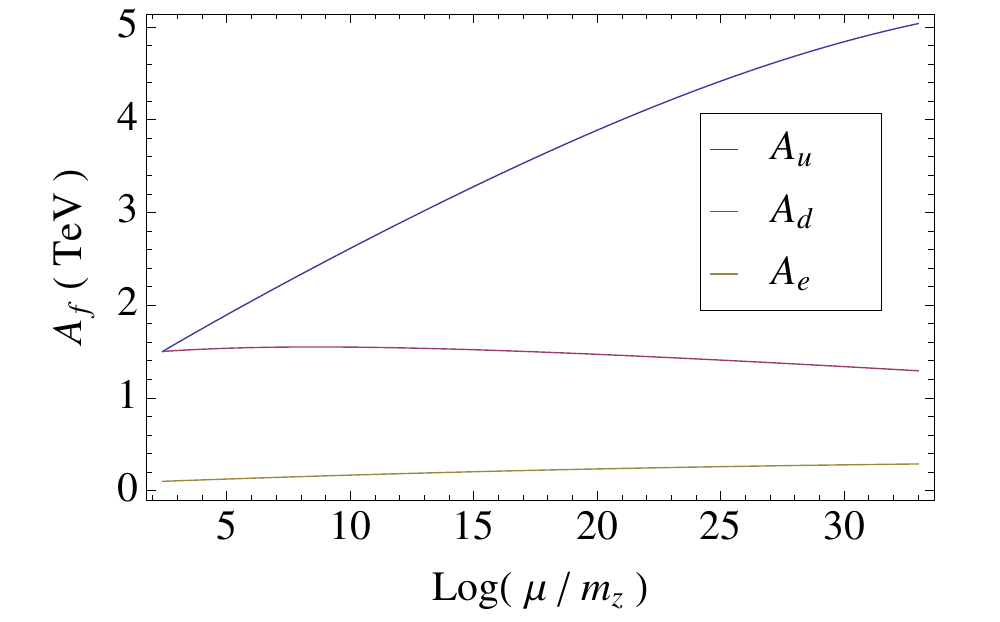}  &
\includegraphics[width=0.45\textwidth]{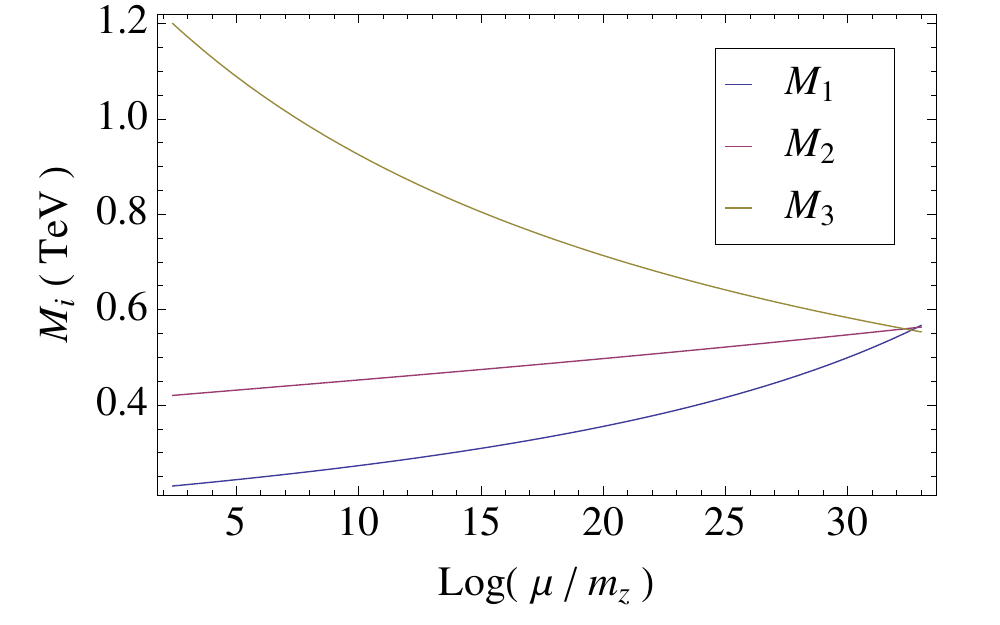}  \\
\end{tabular}
\end{center}
\caption{Mass parameter evolution from the TeV to the Messenger scale $M \simeq  10^{16}$~GeV and $\tan\beta=30$. }
\label{runningteq3330}
\end{figure}

We  present here the RG evolution of the soft supersymmetry-breaking parameters that lead to flavor independent mass parameters at the messenger scale for the light stau scenario. The running of all the soft parameters for the three cases discussed in Sec.~\ref{RGE}, are shown Figs.~\ref{runningteq95},~\ref{runningteq3360},~\ref{runningteq3330}.

As stressed in the text, we have not performed a detailed scanning of the parameters consistent with a 125~GeV Higgs boson with an enhanced diphoton decay rate. We are interested in showing the qualitative behavior of the running of the soft supersymmetry-breaking masses in the region of parameters under study. For this analysis, the threshold corrections to the bottom and tau Yukawa couplings~\cite{deltamb}~\cite{deltamb1}~\cite{deltamb2} have been neglected, since in a bottom-up approach they depend strongly on parameters which are not fixed in a direct way by the Higgs sector. The gluino mass is kept at 1.2 TeV, while $M_2 \simeq 400$~GeV and $M_1 \simeq 200$~GeV. Variations of $M_1$ lead to only a small modification of the RG running of the other parameters. The gluino mass, however, has a strong impact on the running and also modifies the threshold corrections to the Yukwas in a strong way.

We see that values of the messenger scale close to the GUT scale imply boundary conditions for the squark and slepton mass parameters of a few TeV for the example given in Fig.~\ref{runningmL} (a). One interesting effect is that one can obtain a large hierarchy between the third generation and first and second generation mass parameters, with low energy values for the third generation slepton and squark mass parameters that are of the order of the weak scale.  This hierarchy of masses would be induced by the running and would not signal a breakdown of flavor universality at large energies~\cite{Bagger:1999sy},\cite{Bagger:1999ty}.

The gaugino masses at the messenger scale may be an order of magnitude smaller than the scalar masses. Large values of the third generation Yukawa couplings also lead to the interesting property that 
the value of the stau mixing parameter, $A_{\tau}$, tends to be driven to small values. On the other hand, for the relatively large values of the top and bottom Yukawa couplings that are obtained for $\tan\beta \simeq 60$, the large values of $A_{t}$ that are necessary to obtain a 125~GeV Higgs require large values at the messenger scale~\cite{Draper:2011aa}.

\newpage


\begin{thebibliography}{999}
\bibitem{ATLAS:2012ae} 
  G.~Aad {\it et al.}  [ATLAS Collaboration],
  Phys.\ Lett.\ B {\bf 710}, 49 (2012)
  [arXiv:1202.1408 [hep-ex]].
  
   \bibitem{AtlasTwiki}
Atlas Twiki, \\
\href{https://twiki.cern.ch/twiki/bin/view/AtlasPublic/Publications}{https://twiki.cern.ch/twiki/bin/view/AtlasPublic/Publications
}

\bibitem{Chatrchyan:2012tx} 
  S.~Chatrchyan {\it et al.}  [CMS Collaboration],
  arXiv:1202.1488 [hep-ex].
  
     \bibitem{CMSTwiki}
CMS Twiki, \\
\href{https://twiki.cern.ch/twiki/bin/view/CMSPublic/PhysicsResultsHIG}{https://twiki.cern.ch/twiki/bin/view/CMSPublic/PhysicsResultsHIG
}

\bibitem{ALEPH:2010aa} 
  [ALEPH and CDF and D0 and DELPHI and L3 and OPAL and SLD and LEP Electroweak Working Group and Tevatron Electroweak Working Group and SLD Electroweak and Heavy Flavour Groups Collaborations],
  arXiv:1012.2367 [hep-ex].

\bibitem{TevatronElectroweakWorkingGroup:2012gb} 
  Tevatron Electroweak Working Group, f.~t.~C.~Collaboration and D.~Collaboration,
  arXiv:1204.0042 [hep-ex].


\bibitem{LopesdeSa:2012ak}
  R.~Lopes de Sa [CDF and D0 Collaboration],
  arXiv:1204.3260 [hep-ex].
  
\bibitem{TEVNPH:2012ab} 
  [TEVNPH (Tevatron New Phenomina and Higgs Working Group) and CDF and D0 Collaborations],
  arXiv:1203.3774 [hep-ex].
  
  
\bibitem{reviews}
H.~P.~Nilles,
Phys.\ Rept.\  {\bf 110} (1984) 1;\\
H.~E.~Haber and G.~L.~Kane, 
Phys.\ Rept.\  {\bf 117}(1985) 75; \\
S.~P.~Martin,
arXiv:hep-ph/9709356. 


\bibitem{Higgs:1964pi}
  P.~W.~Higgs,
  Phys.\ Rev.\ Lett.\  {\bf 13}, 508-509 (1964).

\bibitem{Higgs:1966ev}
  P.~W.~Higgs,
  Phys.\ Rev.\  {\bf 145}, 1156-1163 (1966).

\bibitem{Okada:1990vk}
  Y.~Okada, M.~Yamaguchi and T.~Yanagida,
  Prog.\ Theor.\ Phys.\  {\bf 85}, 1 (1991).

\bibitem{Ellis:1990nz}
  J.~R.~Ellis, G.~Ridolfi and F.~Zwirner,
  Phys.\ Lett.\  B {\bf 257}, 83 (1991).

\bibitem{Haber:1990aw}
  H.~E.~Haber and R.~Hempfling,
  Phys.\ Rev.\ Lett.\  {\bf 66}, 1815 (1991).

\bibitem{mhiggsRG1a}
  J.~A.~Casas, J.~R.~Espinosa, M.~Quiros and A.~Riotto,
  Nucl.\ Phys.\  B {\bf 436}, 3 (1995)
  [Erratum-ibid.\  B {\bf 439}, 466 (1995)]
  [arXiv:hep-ph/9407389].

\bibitem{mhiggsRG1} M.~Carena, J.~Espinosa, M.~Quir\'os and C.~Wagner,
                    {\em Phys. Lett.} {\bf B 355} (1995) 209,
                    hep-ph/9504316;\\
                    M.~Carena, M.~Quir\'os and C.~Wagner,
                    {\em Nucl. Phys.} {\bf B 461} (1996) 407,
                    hep-ph/9508343.
\bibitem{HHH} H.~Haber, R.~Hempfling and A.~Hoang,
              {\em Z. Phys.} {\bf C 75} (1997) 539,
              hep-ph/9609331.

\bibitem{Heinemeyer:1998yj}
  S.~Heinemeyer, W.~Hollik and G.~Weiglein,
  Comput.\ Phys.\ Commun.\  {\bf 124}, 76 (2000)
  [hep-ph/9812320].
\bibitem{Heinemeyer:1998np}
  S.~Heinemeyer, W.~Hollik and G.~Weiglein,
  Eur.\ Phys.\ J.\ C {\bf 9}, 343 (1999)
  [hep-ph/9812472].
  \bibitem{Carena:2000dp}
  M.~S.~Carena, H.~E.~Haber, S.~Heinemeyer, W.~Hollik, C.~E.~M.~Wagner and G.~Weiglein,
  Nucl.\ Phys.\ B {\bf 580}, 29 (2000)
  [hep-ph/0001002].

\bibitem{Martin:2002wn}
  S.~P.~Martin,
  Phys.\ Rev.\ D {\bf 67}, 095012 (2003)
  [hep-ph/0211366].

\bibitem{Degrassi:2002fi}
  G.~Degrassi, S.~Heinemeyer, W.~Hollik, P.~Slavich and G.~Weiglein,
  Eur.\ Phys.\ J.\ C {\bf 28}, 133 (2003)
  [hep-ph/0212020].


\bibitem{Arbey:2011ab}
  A.~Arbey, M.~Battaglia, A.~Djouadi, F.~Mahmoudi and J.~Quevillon,
  Phys.\ Lett.\  B {\bf 708}, 162 (2012)
  [arXiv:1112.3028 [hep-ph]].

\bibitem{Hall:2011aa}
  L.~J.~Hall, D.~Pinner and J.~T.~Ruderman,
  arXiv:1112.2703 [hep-ph].

\bibitem{Heinemeyer:2011aa}
  S.~Heinemeyer, O.~Stal and G.~Weiglein,
  Phys.\ Lett.\  B {\bf 710}, 201 (2012)
  [arXiv:1112.3026 [hep-ph]].

\bibitem{Feng:2011aa}
  J.~L.~Feng, K.~T.~Matchev and D.~Sanford,
  Phys.\ Rev.\  D {\bf 85}, 075007 (2012)
  [arXiv:1112.3021 [hep-ph]].


\bibitem{Carena:2011aa}
  M.~Carena, S.~Gori, N.~R.~Shah and C.~E.~M.~Wagner,
  JHEP {\bf 1203}, 014 (2012)
  [arXiv:1112.3336 [hep-ph]].

\bibitem{Draper:2011aa}
  P.~Draper, P.~Meade, M.~Reece and D.~Shih,
  arXiv:1112.3068 [hep-ph].



\bibitem{Ellwanger:2011aa}
  U.~Ellwanger,
  JHEP {\bf 1203}, 044 (2012)
  [arXiv:1112.3548 [hep-ph]].

\bibitem{Kane:2011kj}
  G.~Kane, P.~Kumar, R.~Lu and B.~Zheng,
  arXiv:1112.1059 [hep-ph].



\bibitem{Akula:2011aa}
  S.~Akula, B.~Altunkaynak, D.~Feldman, P.~Nath and G.~Peim,
  Phys.\ Rev.\  D {\bf 85}, 075001 (2012)
  [arXiv:1112.3645 [hep-ph]].





\bibitem{Baer:2011ab}
  H.~Baer, V.~Barger and A.~Mustafayev,
  Phys.\ Rev.\  D {\bf 85}, 075010 (2012)
  [arXiv:1112.3017 [hep-ph]].

\bibitem{Kadastik:2011aa}
  M.~Kadastik, K.~Kannike, A.~Racioppi and M.~Raidal,
  arXiv:1112.3647 [hep-ph].

\bibitem{Gunion:2012zd}
  J.~F.~Gunion, Y.~Jiang and S.~Kraml,
  Phys.\ Lett.\  B {\bf 710}, 454 (2012)
  [arXiv:1201.0982 [hep-ph]].

\bibitem{King:2012is}
  S.~F.~King, M.~Muhlleitner and R.~Nevzorov,
  Nucl.\ Phys.\  B {\bf 860}, 207 (2012)
  [arXiv:1201.2671 [hep-ph]].


\bibitem{Cao:2012fz}
  J.~Cao, Z.~Heng, J.~M.~Yang, Y.~Zhang and J.~Zhu,
  JHEP {\bf 1203}, 086 (2012)
  [arXiv:1202.5821 [hep-ph]].
  
\bibitem{Aparicio:2012iw}
  L.~Aparicio, D.~G.~Cerdeno and L.~E.~Ibanez,
  arXiv:1202.0822 [hep-ph].

\bibitem{Ajaib:2012vc}
  M.~A.~Ajaib, I.~Gogoladze, F.~Nasir and Q.~Shafi,
  arXiv:1204.2856 [hep-ph].

\bibitem{Christensen:2012ei}
  N.~D.~Christensen, T.~Han and S.~Su,
  arXiv:1203.3207 [hep-ph].

\bibitem{Brummer:2012ns}
  F.~Brummer, S.~Kraml and S.~Kulkarni,
  arXiv:1204.5977 [hep-ph].

\bibitem{LEPlimit} 
  A.~Heister {\it et al.}  [ALEPH Collaboration],
  Phys.\ Lett.\ B {\bf 526}, 206 (2002)
  [hep-ex/0112011].


\bibitem{Spira:1995rr}
  M.~Spira, A.~Djouadi, D.~Graudenz and P.~M.~Zerwas,
  Nucl.\ Phys.\  B {\bf 453} (1995) 17
  [arXiv:hep-ph/9504378].

  \bibitem{Maltoni}
F. Maltoni et al. (SM Higgs production cross-sections), \\
\href{http://maltoni.home.cern.ch/maltoni/TeV4LHC/}{
http://maltoni.home.cern.ch/maltoni/TeV4LHC/
}


  
  
\bibitem{Anastasiou:2008tj} 
  C.~Anastasiou, R.~Boughezal and F.~Petriello,
  JHEP {\bf 0904}, 003 (2009)
  [arXiv:0811.3458 [hep-ph]].


\bibitem{deFlorian:2009hc} 
  D.~de Florian and M.~Grazzini,
  Phys.\ Lett.\ B {\bf 674}, 291 (2009)
  [arXiv:0901.2427 [hep-ph]].


  \bibitem{Grazzini}
M. Grazzini et al. (SM ggH prediction), \\
\href{http://theory.fi.infn.it/grazzini/hcalculators.html}{
http://theory.fi.infn.it/grazzini/hcalculators.html
}

\bibitem{Djouadi:1998az}
  A.~Djouadi,
  Phys.\ Lett.\ B {\bf 435}, 101 (1998)
  [hep-ph/9806315].
  
\bibitem{Dermisek:2007fi} 
  R.~Dermisek and I.~Low,
  Phys.\ Rev.\ D {\bf 77}, 035012 (2008)
  [hep-ph/0701235 [HEP-PH]].
  
  \bibitem{FH}
Home of FeynHiggs, \\
\href{http://wwwth.mpp.mpg.de/members/heinemey/feynhiggs/cFeynHiggs.html}{
http://wwwth.mpp.mpg.de/members/heinemey/feynhiggs/cFeynHiggs.html
}

\bibitem{Frank:2006yh}
  M.~Frank, T.~Hahn, S.~Heinemeyer, W.~Hollik, H.~Rzehak and G.~Weiglein,
  JHEP {\bf 0702}, 047 (2007)
  [hep-ph/0611326].
  
\bibitem{CPsuperH}
  J.~S.~Lee, A.~Pilaftsis, M.~S.~Carena, S.~Y.~Choi, M.~Drees, J.~R.~Ellis and C.~E.~M.~Wagner,
  Comput.\ Phys.\ Commun.\  {\bf 156}, 283 (2004)
  [hep-ph/0307377]; \\
 J.~S.~Lee, M.~Carena, J.~Ellis, A.~Pilaftsis and C.~E.~M.~Wagner,
  Comput.\ Phys.\ Commun.\  {\bf 180}, 312 (2009)
  [arXiv:0712.2360 [hep-ph]].

  
\bibitem{Diaz:2004qt}
  M.~A.~Diaz and P.~Fileviez Perez,
  J.\ Phys.\ G {\bf 31}, 563 (2005)
  [arXiv:hep-ph/0412066].
  
  \bibitem{Ellis:1975ap} 
  J.~R.~Ellis, M.~K.~Gaillard and D.~V.~Nanopoulos,
  Nucl.\ Phys.\ B {\bf 106}, 292 (1976).


\bibitem{Shifman:1979eb}
  M.~A.~Shifman, A.~I.~Vainshtein, M.~B.~Voloshin and V.~I.~Zakharov,
  Sov.\ J.\ Nucl.\ Phys.\  {\bf 30}, 711 (1979)
  [Yad.\ Fiz.\  {\bf 30}, 1368 (1979)].
  
  
  
\bibitem{Aad:2011rv}  
ÊG.~Aad {\it et al.} Ê[ATLAS Collaboration], Ê
Phys.\ Lett.\ B {\bf 705}, 174 (2011) Ê[arXiv:1107.5003 [hep-ex]]. Ê

  \bibitem{Atlastau}
Atlas $\tau$ Public Note, \\
\href{http://cdsweb.cern.ch/record/1429662
}{http://cdsweb.cern.ch/record/1429662
}



\bibitem{Chatrchyan:2012vp}  Ê
S.~Chatrchyan {\it et al.} Ê[CMS Collaboration], 
Ê
ÊarXiv:1202.4083 [hep-ex]. Ê

  
     \bibitem{CMStau}
CMS $\tau$ Public Note, \\
\href{http://cdsweb.cern.ch/record/1429929/files/HIG-12-007-pas.pdf}{http://cdsweb.cern.ch/record/1429929/files/HIG-12-007-pas.pdf
}

\bibitem{Carena:2012rw}  ÊM.~Carena, S.~Gori, A.~Juste, A.~Menon, C.~E.~M.~Wagner and L.~-T.~Wang, Ê
arXiv:1203.1041 [hep-ph]. Ê


\bibitem{arXiv:1107.4354}
  M.~Carena, P.~Draper, T.~Liu and C.~E.~M.~Wagner,
  Phys.\ Rev.\ D\ {\bf 84}, 095010  (2011)
  [arXiv:1107.4354 [hep-ph]].
 


\bibitem{Hisano:2010re} 
  J.~Hisano and S.~Sugiyama,
  Phys.\ Lett.\ B {\bf 696}, 92 (2011)
  [arXiv:1011.0260 [hep-ph]].

\bibitem{Chankowski:1993eu}
  P.~H.~Chankowski, A.~Dabelstein, W.~Hollik, W.~M.~Mosle, S.~Pokorski and J.~Rosiek,
  Nucl.\ Phys.\  B {\bf 417}, 101 (1994).

\bibitem{Heinemeyer:2006px} 
  S.~Heinemeyer, W.~Hollik, D.~Stockinger, A.~M.~Weber and G.~Weiglein,
  JHEP {\bf 0608}, 052 (2006)
  [hep-ph/0604147].
  







%








\bibitem{Heinemeyer:2004gx} 
  S.~Heinemeyer, W.~Hollik and G.~Weiglein,
  Phys.\ Rept.\  {\bf 425}, 265 (2006)
  [hep-ph/0412214].
  
 
  
\bibitem{Moroi:1995yh}
  T.~Moroi,
  Phys.\ Rev.\  D {\bf 53}, 6565 (1996)
  [Erratum-ibid.\  D {\bf 56}, 4424 (1997)]
  [arXiv:hep-ph/9512396].

\bibitem{Chattopadhyay:1995ae} 
  U.~Chattopadhyay and P.~Nath,
  Phys.\ Rev.\ D {\bf 53}, 1648 (1996)
  [hep-ph/9507386].

\bibitem{Carena:1996qa} 
  M.~S.~Carena, G.~F.~Giudice and C.~E.~M.~Wagner,
  Phys.\ Lett.\ B {\bf 390}, 234 (1997)
  [hep-ph/9610233].
  
\bibitem{Martin:2001st} 
  S.~P.~Martin and J.~D.~Wells,
  Phys.\ Rev.\ D {\bf 64}, 035003 (2001)
  [hep-ph/0103067].
 
\bibitem{Feng:2001tr} 
  J.~L.~Feng and K.~T.~Matchev,
  Phys.\ Rev.\ Lett.\  {\bf 86}, 3480 (2001)
  [hep-ph/0102146].
  
\bibitem{Jegerlehner:2009ry} 
  F.~Jegerlehner and A.~Nyffeler,
  Phys.\ Rept.\  {\bf 477}, 1 (2009)
  [arXiv:0902.3360 [hep-ph]].
  
\bibitem{Nakamura:2010zzi} 
  K.~Nakamura {\it et al.}  [Particle Data Group Collaboration],
  J.\ Phys.\ G G {\bf 37}, 075021 (2010) and 2011 partial update for the 2012 edition.
  
\bibitem{Edsjo:1997bg} 
  J.~Edsjo and P.~Gondolo,
  Phys.\ Rev.\ D {\bf 56}, 1879 (1997)
  [hep-ph/9704361].
  
\bibitem{Ellis:1999mm} 
  J.~R.~Ellis, T.~Falk, K.~A.~Olive and M.~Srednicki,
  Astropart.\ Phys.\  {\bf 13}, 181 (2000)
  [Erratum-ibid.\  {\bf 15}, 413 (2001)]
  [hep-ph/9905481].
  
\bibitem{Belanger:2010pz} 
  G.~Belanger, F.~Boudjema, A.~Pukhov and A.~Semenov,
  arXiv:1005.4133 [hep-ph].
  
\bibitem{Gondolo:2004sc} 
  P.~Gondolo, J.~Edsjo, P.~Ullio, L.~Bergstrom, M.~Schelke and E.~A.~Baltz,
  JCAP {\bf 0407}, 008 (2004)
  [astro-ph/0406204].
  


\bibitem{Martin:1993zk} 
  S.~P.~Martin and M.~T.~Vaughn,
  Phys.\ Rev.\ D {\bf 50}, 2282 (1994)
  [Erratum-ibid.\ D {\bf 78}, 039903 (2008)]
  [hep-ph/9311340].
  
\bibitem{Carena:1993bs} 
  M.~S.~Carena, M.~Olechowski, S.~Pokorski and C.~E.~M.~Wagner,
  Nucl.\ Phys.\ B {\bf 419}, 213 (1994)
  [hep-ph/9311222].
  
\bibitem{Carena:1994bv} 
  M.~S.~Carena, M.~Olechowski, S.~Pokorski and C.~E.~M.~Wagner,
  Nucl.\ Phys.\ B {\bf 426}, 269 (1994)
  [hep-ph/9402253].
  
\bibitem{Bagger:1999sy} 
  J.~A.~Bagger, J.~L.~Feng, N.~Polonsky and R.~-J.~Zhang,
  Phys.\ Lett.\ B {\bf 473}, 264 (2000)
  [hep-ph/9911255].
  

\bibitem{Bagger:1999ty} 
  J.~Bagger, J.~L.~Feng and N.~Polonsky,
  Nucl.\ Phys.\ B {\bf 563}, 3 (1999)
  [hep-ph/9905292].
  

\bibitem{Blazek:2002ta} 
  T.~Blazek, R.~Dermisek and S.~Raby,
  Phys.\ Rev.\ D {\bf 65}, 115004 (2002)
  [hep-ph/0201081].
  
  
  \bibitem{deltamb}
 L. Hall, R. Rattazzi and U. Sarid,  Phys. Rev.
D{\bf50} (1994) 7048; \\
R. Hempfling, Phys. Rev. D{\bf49} (1994) 6168.
\bibitem{deltamb1} 
M. Carena, M. Olechowski, S. Pokorski and C.E.M. Wagner,
Nucl. Phys. B{\bf 426} (1994) 269.
\bibitem{deltamb2}  D. Pierce, J. Bagger, K. Matchev, and R. Zhang,
Nucl. Phys. {B}{\bf 491} (1997) 3.







 















\bibitem{Alwall:2011uj} 
  J.~Alwall, M.~Herquet, F.~Maltoni, O.~Mattelaer and T.~Stelzer,
  JHEP {\bf 1106}, 128 (2011)
  [arXiv:1106.0522 [hep-ph]].



\bibitem{Aad:2012rt} 
  G.~Aad {\it et al.}  [ATLAS Collaboration],
  arXiv:1204.3852 [hep-ex].
    
\bibitem{ATLAS:2012ag} 
  G.~Aad {\it et al.}  [ATLAS Collaboration],
  arXiv:1203.6580 [hep-ex].
\bibitem{Chatrchyan:2012ye} 
  S.~Chatrchyan {\it et al.}  [CMS Collaboration],
  arXiv:1204.5341 [hep-ex].
  
  
  \bibitem{ATLAS:tau}
 ATLAS collaboration,
  ATLAS-CONF-2011-152.
  
      \bibitem{ATLAS:ditau}
 ATLAS collaboration,
  ATLAS-CONF-2012-014.
  
  
\bibitem{Cao:2003tr} 
  Q.~-H.~Cao, S.~Kanemura and C.~P.~Yuan,
  Phys.\ Rev.\ D {\bf 69}, 075008 (2004)
  [hep-ph/0311083].
  
\bibitem{Lindert:2011td} 
  J.~M.~Lindert, F.~D.~Steffen and M.~K.~Trenkel,
  JHEP {\bf 1108}, 151 (2011)
  [arXiv:1106.4005 [hep-ph]].






\end{thebibliography}
\end{document}